\documentclass[preprint,onecolumn,nofootinbib]{revtex4}
\pdfoutput=1
\usepackage[colorlinks=true,linkcolor=blue,urlcolor=blue,filecolor=black,citecolor=red,pdfstartview=FitV,pdftitle={},pdfsubject={},pdfkeywords={},pdfpagemode=None,bookmarksopen=true]{hyperref}
\usepackage{graphicx}
\usepackage{amsmath}
\usepackage{amsfonts}
\usepackage{amssymb,ulem}
\usepackage{color,xcolor}%
\usepackage{CJK}
\usepackage{subfigure}

\usepackage{dcolumn}
\newcommand{\f}{\begin{equation}}
    \newcommand{\ff}{\end{equation}}
\newcommand{\fa}{\begin{eqnarray}}
    \newcommand{\ffa}{\end{eqnarray}}
\newcommand{\bsub}{\begin{subequations}}
    \newcommand{\esub}{\end{subequations}}

\begin{document}
\title{Alternating current conductivity and superconducting properties of the holographic effective theory}

\author{Yan Liu$^{1}$}
\author{Xi-Jing Wang$^{2}$}
\author{Jian-Pin Wu$^{2}$}
\thanks{jianpinwu@yzu.edu.cn}
\author{Xin Zhang$^{1,3,4}$}
\thanks{zhangxin@mail.neu.edu.cn}
\affiliation{
    $^1$ Department of Physics, College of Sciences, Northeastern University, Shenyang 110819, China \\
    $^2$ Center for Gravitation and Cosmology, College of Physical Science and Technology, Yangzhou University, Yangzhou 225009, China \\
     $^3$ Frontiers Science Center for Industrial Intelligence and Systems Optimization, Northeastern University, Shenyang 110819, China \\
    $^4$  Key Laboratory of Data Analytics and Optimization for Smart Industry, Ministry of Education, Northeastern University, Shenyang 110819, China
}

\begin{abstract}
    We construct a holographic effective superconducting theory by considering a special gauge-axion higher derivative term. The gauge-axion coupling results in the transport behavior similar to the vortex response in the dual boundary field theory leading to non-Drude behavior of alternating current (AC) conductivity at the weak momentum dissipation. With the momentum dissipation increasing, a dip exhibits in the AC conductivity at low frequency. It is thought to be the result of a combination of the strong momentum dissipation and the gauge-axion coupling. In the superconducting phase, this gauge-axion coupling also plays a key role leading to a more evident gap at the low frequency conductivity. In addition, we also study the combined effects of the strength of momentum dissipation and various couplings among the gauge field, axion fields and the complex scalar field.

\end{abstract}

\maketitle
\tableofcontents

\section{Introduction}
The Bardeen-Cooper-Schrieffer (BCS) theory \cite{Bardeen:1957kj,Bardeen:1957mv} is a successful microscopic theory to describe the superconductivity discovered in 1911 \cite{Onnes}. However, because the Cooper pairs decouple and no longer exist as the temperature of system increases, the BCS theory loses its power for the high temperature superconductors. Many scientists have been trying to reveal the basic principle behind them. But until now it still has no satisfied microscopic description for the high temperature superconductivity.

Alternatively, AdS/CFT correspondence \cite{Maldacena:1997re,Gubser:1998bc,Witten:1998qj,Aharony:1999ti}, or more generally referring to holography, provides an effective description for the high temperature superconductors and also opens up a new window for studying the mechanism of the high temperature superconductivity. The main idea of holography is that a strong coupled quantum field theory can be mapped to a higher-dimensional bulk gravity system. Based on this idea, a pioneering holographic superconductor model is proposed by Hartnoll, Herzog, and Horowitz (HHH) \cite{Hartnoll:2008vx,Hartnoll:2008kx}. This model exists two states: the superconducting state with a non-vanishing charge condensate and the normal state of a perfect conductor. It means that there exists an infinite electric infinite direct current (DC) conductivity even in the normal state due to the translational invariance of the system. To construct a more realistic superconductor model, it is important to break the translational invariance in the holographic framework such that we have a momentum dissipation system.

There are several ways to break the translational invariance in the bulk gravity. The most brutal way is to introduce a spatially-dependent source in the dual boundary theory. It has been implemented by a spatially-periodic real scalar source or chemical potential \cite{Horowitz:2012ky,Horowitz:2012gs,Ling:2013aya,Ling:2013nxa,Donos:2014yya}. They are usually referred to as the scalar lattice and the ionic lattice, respectively. In this method, such spatially-dependent sources typically lead to the inhomogeneity of the dynamic fields in the bulk, which generate a set of complicated coupled partial differential equations (PDEs). Though conceptually clear, the application of this framework is limited by numerical technology. This is because the accuracy of solving PDE numerically depends heavily on the temperature of the background. Thus, it is extremely difficult to explore the lattice effects at the extremely low temperature~\cite{Hartnoll:2014gaa}.

To bypass the technical complexity of solving the PDEs but capture the important aspect of the momentum dissipation, people develop a much simpler but elegant mechanism to break the translation symmetry but retain the homogeneity of the background geometry, for which we only need solve the ordinary differential equations (ODEs). So far, several models have been proposed to implement this mechanism, including holographic Q-lattices \cite{Donos:2013eha,Donos:2014uba,Ling:2015epa,Ling:2015exa}, helical lattices \cite{Donos:2012js} and axions model \cite{Andrade:2013gsa,Kim:2014bza,Cheng:2014tya,Ge:2014aza,Andrade:2016tbr,Kuang:2017cgt,Kuang:2016edj,Tanhayi:2016uui,Cisterna:2017jmv,Cisterna:2017qrb,Cisterna:2018hzf}.
Holographic Q-lattice model implements the breaking of the translational symmetry by the global phase of the complex scalar field. Holographic helical lattice model possesses the non-Abelian Bianchi VII$_0$ symmetry, which results in the translational symmetry breaking but holds homogeneous background geometry. The simplest model is the so-called holographic axions model \cite{Andrade:2013gsa,Baggioli:2021xuv}, which breaks the translational symmetry by a pair of linearly spatial-dependent scalar fields. Further, guiding in the spirit of effective holographic low energy theories, it is natural and interesting to explore the effect of higher-derivative terms of axion fields \cite{Baggioli:2016pia,Huh:2021ppg,Gouteraux:2016wxj,Li:2018vrz}. The higher-derivative effect turns out to have a heavy impact on the lower bound of charge diffusion \cite{Baggioli:2016pia} but have no effect on its upper bound \cite{Huh:2021ppg}. Such higher-derivative term has no impact on the bound of energy diffusion \cite{Baggioli:2016pia}. Further, it was found that the lower bound on the DC conductivity in the usual axions model \cite{Andrade:2013gsa} is violated such that we have vanishing DC conductivity at zero temperature, which provides a framework to model a more realistic insulating state.

Massive gravity provides a alternative mechanism to implement the momentum dissipation. In massive gravity theory the diffeomorphism invariance in the bulk is broken and ultimately results in the broken translational invariance in the dual boundary field theory \cite{Vegh:2013sk,Davison:2013jba,Blake:2013bqa}.
In addition, we can also achieve the momentum dissipation in the dual boundary field theory by introducing a higher-derivative interaction term between the $U(1)$ gauge field and the scalar field, which spontaneously generates a spatially dependent profile of the scalar field \cite{Kuang:2013oqa,Alsup:2013kda}.

The holographic superconductor models with momentum dissipation have also been constructed and studied \cite{Horowitz:2013jaa,Ling:2014laa,Ling:2017naw,Kim:2015dna,Jeong:2021wiu,Andrade:2014xca,Zeng:2014uoa,Baggioli:2015dwa}. In Ref.~\cite{Horowitz:2013jaa}, a pioneering work of holographic lattice superconductor model implemented by a periodic potential was set up. They reproduced some qualitative features of some cuprates, including the superconducting energy gap and the power law fall-off at the intermediate frequency. The holographic Q-lattice superconductor was also built in \cite{Ling:2014laa,Ling:2017naw}. Their result demonstrates that the condensate of the scalar field would be suppressed by the lattice effects and thus leads to a lower critical temperature. In particular, they found that if the normal state is a deep insulating phase, the condensation never happens for small charge of the scalar field. Moreover in holographic superconductor with axion fields \cite{Kim:2015dna}, the authors found the existence of a new type of superconductor induced by the strength of momentum relaxation even at chemical potential being zero. It means that there exists a new ``pairing'' mechanism of particles and antiparticles interacting with the strength of momentum relaxation. In Ref. \cite{Jeong:2021wiu}, a more phenomenologically relevant holographic superconducting model of momentum dissipation was constructed based on Gubser-Rocha model whose ground state entropy vanshes  \cite{Gubser:2009qt}. It was also shown that some universal properties of high $T_c$ superconductors, including linear-T resistivity near $T_c$ and Homes' law, are observed \cite{Jeong:2021wiu}.
In addition, a superconducting dome-shaped region is implemented on the temperature-doping phase diagram in holographic axion superconductor proposed in \cite{Baggioli:2015dwa}.

In this paper, we shall construct an effective holographic superconductor of momentum dissipation, for which the higher derivative terms of axion fields are introduced. We intend to study the effects of higher-derivative couplings on the alternating current (AC) conductivity on the normal state, which is still absent until now, and the superconducting properties.
The plan of this work is as follows: In Sec.~\ref{sec-setup}, we construct the effective holographic model with the higher derivative terms of axion fields. In Sec.~\ref{Normal}, we study the conductivity in the normal state.
The superconducting properties are explored in Sec.~\ref{The Superconducting Phase}. In Sec.~\ref{conclusion}, we conclude.

\section{Holographic framework}\label{sec-setup}
In this section, we construct an effective holographic superconductor model with the momentum dissipation. It includes the following key ingredients: the metric $g_{\mu\nu}$, $U(1)$ gauge field $A_{\mu}$, the complex scalar field $\psi$ and two axionic fields $X^{I}$ ($I=x\,,y$) with $x$, $y$ being spatial coordinates. The superconducting phase is supported by the complex scalar field $\psi$, which can be defined by $\psi = \chi e^{i\theta}$ with $\chi$ being the real scalar field and $\theta$ being the St\"uckelberg field. A pair of spatial linear dependent axionic fields are responsible for the momentum dissipation.

Including all the all ingredients, we write down the total action of the model as
\fa
&&
\label{action}
S=\int d^4x\sqrt{-g}\Big(R+6+\mathcal{L}_{M}+\mathcal{L}_{\chi}+\mathcal{L}_{X}\Big)\,,
\
\\
&&
\label{LM}
\mathcal{L}_{M}=-\frac{Z(\chi)}{4}F^{2}\,,
\
\\
&&
\mathcal{L}_{\chi}=-\frac{1}{2}(\partial_{\mu} \chi)^{2}
-H(\chi)(\partial_{\mu}\theta-q A_{\mu})^{2}-V_{\rm int}(\chi)\,,
\label{Lchi}
\\
\
&&
\mathcal{L}_{X}=-\frac{J(\chi)}{4}{\rm{Tr}}[XF^{2}]-V(X)\,.
\label{lagarangex}
\ffa
$F_{\mu\nu}=\nabla_{\mu}A_{\nu}-\nabla_{\nu}A_{\mu}$ is the field strength of gauge field $A_{\mu}$.
The coupling function of Maxwell field strength takes the form as
\fa
&&
\label{Zchi}
Z(\chi)=1+\frac{a\chi^{2}}{2}\,.
\ffa
We define $X\equiv {\rm{Tr}}[X^{\mu}\,_{\nu}]$ with
\fa
&&
X^{\mu}\,_{\nu}=\frac{1}{2}\partial^{\mu}X^{I}\partial_{\nu}X^{I}\,,
\ffa
where $I=x,y$. The potential $V(X)$ in the Lagrangian density $\mathcal{L}_{X}$ is
\fa
&&
V(X)=X\,,
\ffa
which is just the simplest axion term proposed in Ref.~\cite{Andrade:2013gsa}.
As in Refs.~\cite{Baggioli:2016pia,Gouteraux:2016wxj,Li:2018vrz}, we introduce a higher derivative term of axion fields coupled with the gauge field as \footnote{There is another class of gauge-axion couplings, for instance, $K\,{\rm{Tr}}[X]F^2$ \cite{Gouteraux:2016wxj,Baggioli:2016pia,An:2020tkn,Wang:2021jfu}. This term will change the background solution, but the current model we will consider does not. See below.}
\fa
{\rm{Tr}}[XF^2]\equiv X^{\mu}\,_{\nu}F^{\nu}\,_{\rho}F^{\rho}\,_{\mu}\,.
\ffa
The coupling coefficient $J(\chi)$ takes the form
\fa
\label{Jchi}
J(\chi)=\alpha_{1}+\frac{\alpha_{2}\chi^{2}}{2}\,.
\ffa
$\mathcal{L}_{\chi}$ is the Lagrangian density supporting the superconducting phase transition. The coupling function $H(\chi)$ and the potential $V_{\rm int}(\chi)$ are given by
\fa
H(\chi)=\frac{n\chi^{2}}{2}\,,~~~~~~V_{int}(\chi)=\frac{M^{2}\chi^{2}}{2}\,,
\ffa
where $M$ is the mass of the scalar field $\psi$.
Without loss of generality, we choose the gauge $\theta=0$ in what follows.

From the action above, we derive the covariant form of the equations of motion:
\fa
&&
\label{maxwell-eq}
\nabla_{\mu}\Bigg[ZF^{\mu\nu}-\frac{J}{2}\Big((XF)^{\mu\nu}-(XF)^{\nu\mu}\Big)\Bigg]-2Hq^{2}A^{\mu}=0\,,
\\
&&\label{11}
\nabla_{\mu}\nabla^{\mu}\chi-\partial_{\chi}Hq^{2}A^2-\frac{\partial_{\chi}Z}{4}F^{2}-\frac{\partial_{\chi}J}{4}{\rm{Tr}}[XF^{2}]-\partial_{\chi}V_{int}=0\,,
\\
&&
\label{axion-eq}
\nabla_{\mu}\Big[\nabla^{\mu}X^{I}+\frac{J}{4}(F^{2})^{\mu}\,_{\nu}\nabla^{\nu}X^{I}\Big]=0\,,
\ffa
and
\fa
&&
R_{\mu\nu}-\frac{1}{2}g_{\mu\nu}R-3g_{\mu\nu}-\frac{1}{2}\nabla_{\mu}\chi\nabla_{\nu}\chi-Hq^{2}A_{\mu}A_{\nu}-\frac{Z}{2}F_{\mu\rho}F_{\nu}\,^{\rho}-\frac{J}{4}\Big(\frac{1}{2}\nabla_{(\mu|}X^I\nabla_{\sigma}
\nonumber
\\
&&
X^I(F^2)^{\sigma}\,_{|\nu)}+F_{(\mu|\sigma}(FX)^{\sigma}\,_{|\nu)}+F_{(\mu|\sigma}(XF)^{\sigma}\,_{|\nu)}\Big)-\frac{1}{2}g_{\mu\nu} (\mathcal{L}_{M}+\mathcal{L}_{\chi}+\mathcal{L}_{X})=0\,,
\label{einstein-eq}
\ffa
where the symmetry brackets above mean $A_{(\mu\nu)} = (A_{\mu\nu} + A_{\nu\mu})/2$.

To solve the above equations, we take the following ansatz
\fa
&&
ds^2={1\over u^2}\left[-(1-u)p(u)U_{1}dt^2+\frac{du^2}{(1-u)p(u)U_{1}}+V_{1} dx^2+V_{1} dy^2\right],
\nonumber
\\
&&
A_{t}=\mu(1-u)b(u)\,,\,\,\,\,\chi=u^{3-\Delta}\phi\,,\,\,\,\,
\nonumber
\\
&&
X^{x}=\alpha x\,,\,\,\,\,X^{y}=\alpha y\,,
\label{bhbs}
\ffa
with $p(u)=1+u+u^2-\mu^2u^3/4$ and $\Delta=3/2\pm(9/4+M^2)^{1/2}$.
$u=1$ is the black hole horizon while the AdS boundary locates at $u=0$. $\alpha$ is a constant denoting the strength of momentum dissipation. $U_{1}$ and $V_{1}$ are just the function of the radial coordinate $u$. We impose the boundary condition $U_{1}(1)=1$ at the horizon so that we have the temperature of the dual system as
\fa
\label{tem}
T=\frac{3}{4\pi}-\frac{\mu^2}{16\pi}\,.
\ffa
Through this paper we take $M^2=-2$ such that $\Delta=2$. Then for given coupling parameters $a$, $\alpha_1$, $\alpha_2$ and $n$, this holographic system is depicted by the dimensionless Hawking temperature $T/\mu$ and the strength of momentum dissipation $\alpha/\mu$. For convenience, we abbreviate the two dimensionless quantities $\{T/\mu,\alpha/\mu\}$ to $\{T,\alpha\}$ in the following.

\section{Conductivity in normal phase}\label{Normal}
Obviously, in the case of $\chi=0$, the action describes the non-superconducting phase. The properties of DC transports over the normal phase have been explored in \cite{Li:2018vrz}. But the frequency dependent conductivity of holographic higher-derivative axions model over the normal state is still absent up to now. Therefore, before studying the properties of the superconducting phase, we shall explore the properties of AC conductivity over the normal phase in this section.

In the case of $\chi=0$, the black hole solution can be analytically worked out.
It is found that the gauge-axion coupling has no effect on the background solution and this exact background solution is just the Reissner-Nordstr{\"o}m-AdS (RN-AdS) black hole solution with axions \cite{Andrade:2013gsa}. Here for numerical convenience, we still take the ansatz \eqref{bhbs} to solve the background of the normal state.

Though the gauge-axion coupling has no effect on the background solution, it enters into the perturbative equations leading to strong impact on the transports, which have been seen in the properties of DC transports \cite{Baggioli:2016pia,Gouteraux:2016wxj,Li:2018vrz}. So we expect that AC conductivity in the normal state shall also exhibit some novel properties. For completeness, we shall first give a brief review on DC conductivity in what follows, and then study AC conductivity in normal phase.

\subsection{DC Conductivity}
Based on ``membrane paradigm'' \cite{Iqbal:2008by,Donos:2014cya} (see also \cite{Amoretti:2014mma,Donos:2015gia,Ling:2016dck}), we can analytically work out DC conductivity determined by the background data at the horizon. According to the pivotal point of this method, we construct a radially conserved current $\mathcal{J}$ connecting the horizon and the boundary. At this stage, we can calculate DC conductivity as
\fa
&&
\sigma_{DC}=-\frac{(\alpha^{2}\mu^{2}\alpha_{1}-4V_{1}(1))(\alpha^{2}\mu^{2}+\mu^{2}b^{2}(1)V_{1}(1))}{\alpha^{2}\mu^{2}(4+\alpha_{1}\mu^{2}b^{2}(1))V_{1}(1)}\,.
\label{dc-exp}
\ffa
Obviously, different from the simplest axionic model \cite{Andrade:2013gsa}, DC conductivity from the higher-derivative corrections depends on the temperature $T$ and also the momentum dissipation strength $\alpha$. Notice that the requirement of the positive definiteness of the conductivity imposes a constraint on the coupling parameter $\alpha_1$ as $0\leq\alpha_1\leq2/3$ \cite{Gouteraux:2016wxj}.

\begin{figure}
    \center{
        \includegraphics[scale=0.6]{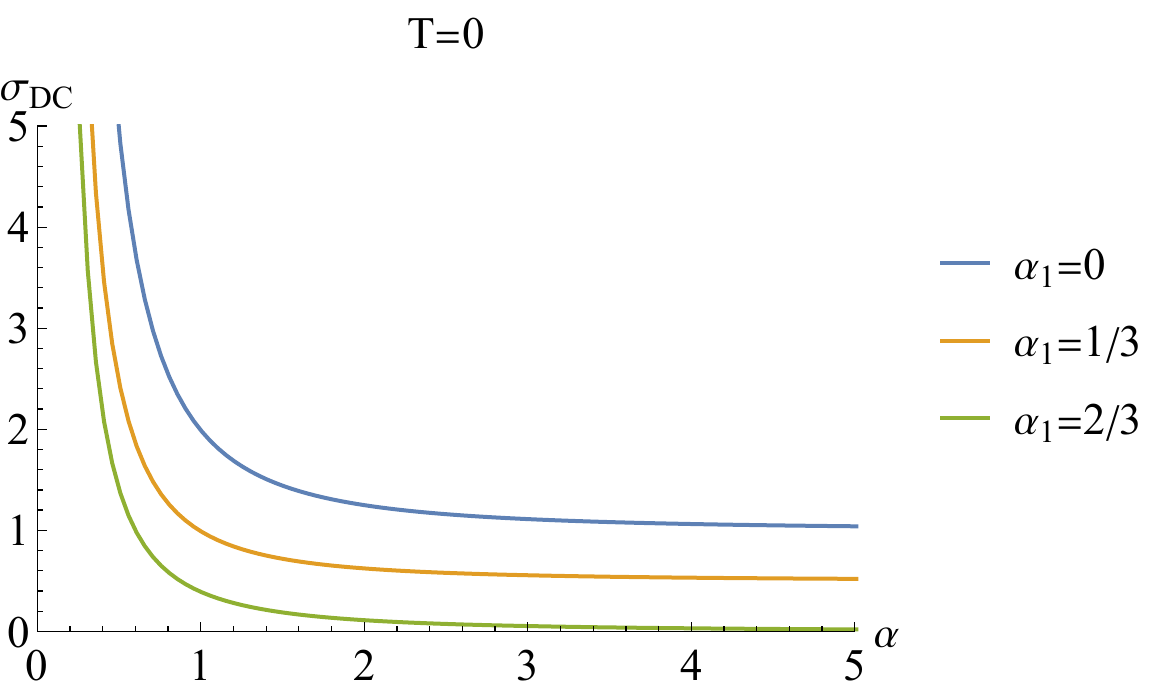}\hspace{0.5cm}
        \includegraphics[scale=0.6]{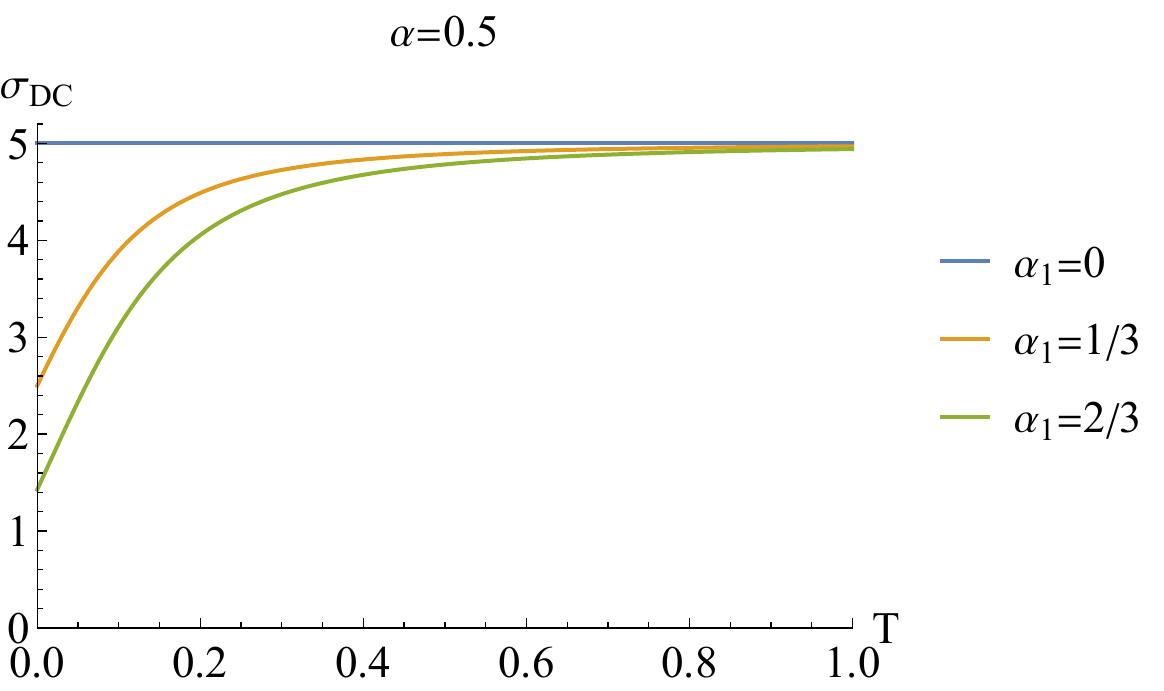}\\
        \includegraphics[scale=0.6]{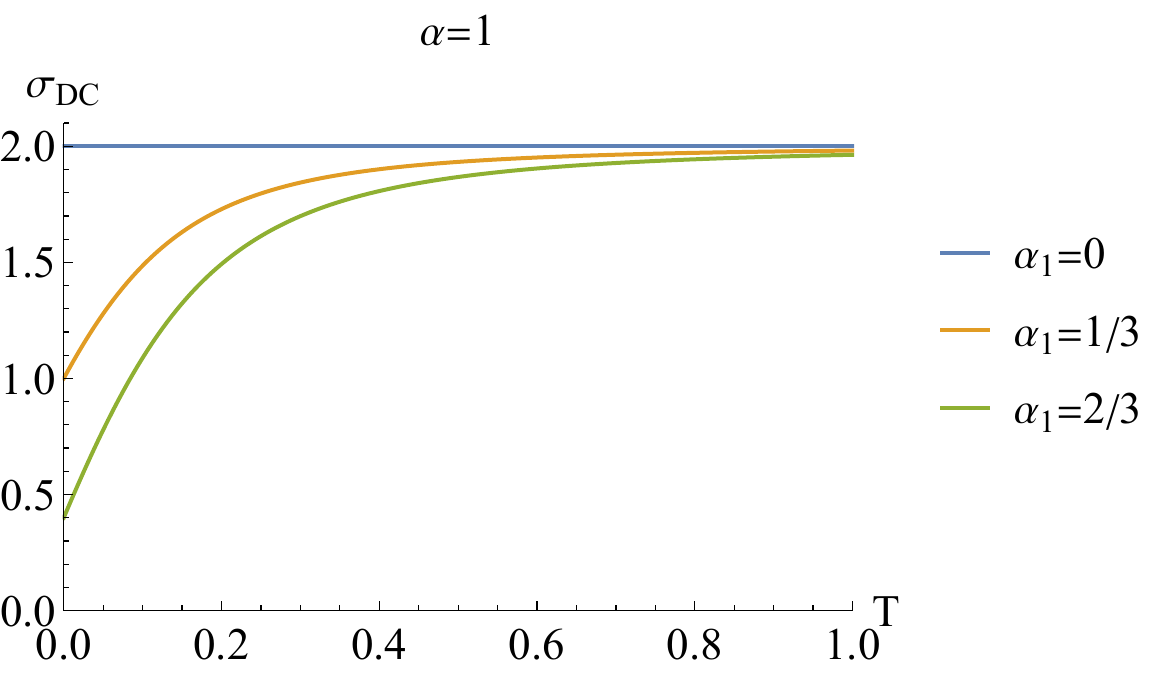}\hspace{0.5cm}
        \includegraphics[scale=0.6]{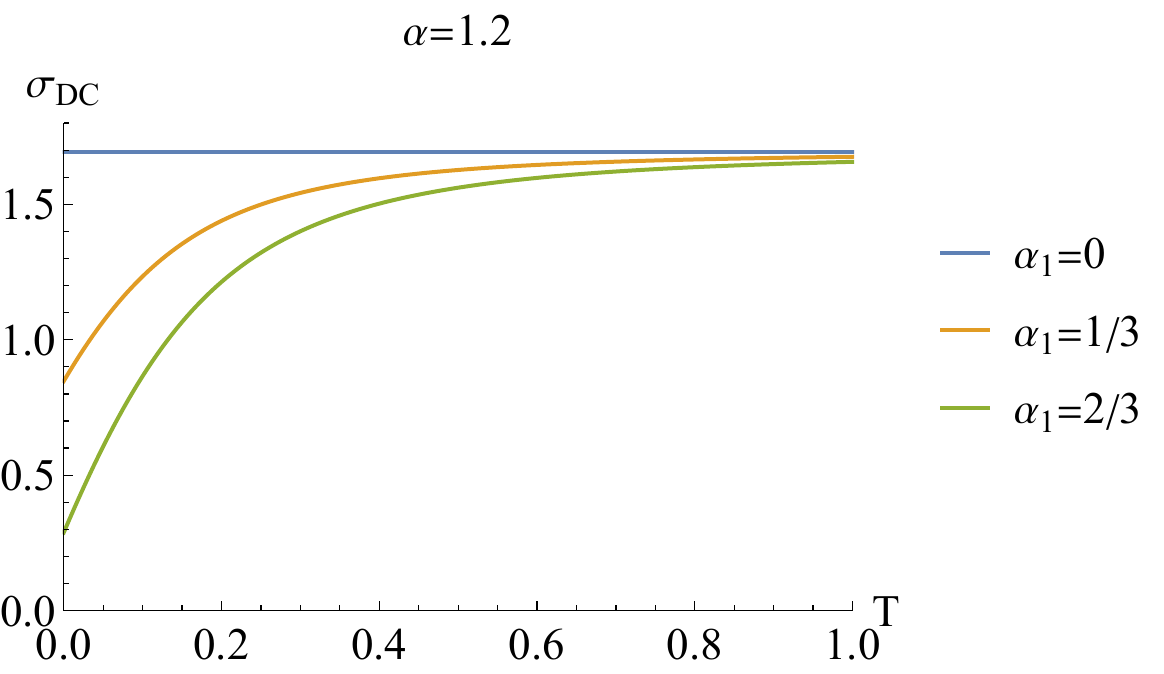}
        \caption{\label{DCcalculate} The DC conductivity with the $\alpha_{1}$ coupling turned on by setting $T=0$ while $\alpha$ vary and by setting $\alpha=\{0.5,1,1.2\}$ while $T$ vary.}}
\end{figure}

We plot DC conductivity with the coupling parameter $\alpha_{1}$ turned on in the top left in Fig.~\ref{DCcalculate}. The plot is exhibited to demonstrate DC conductivity as the function of $\alpha$ at zero temperature. It is easy to find that at zero temperature, when the coupling parameter $\alpha_{1}$ increases, DC conductivity decreases with $\alpha$ increasing. In particular, for $\alpha_{1}=2/3$, the DC conductivity tends to zero with $\alpha$ increasing, which violates the DC conductivity bound in the simplest holographic axion model \cite{Andrade:2013gsa}. Notice that in the limit of $\alpha\rightarrow 0$, due to the restoration of the violation of the translational symmetry, the DC conductivity tends to $\alpha_1$-independent infinity.

In holographic system, the temperature behavior of DC conductivity can well describe some characteristics of the system. In particular, in many holographic references \cite{Baggioli:2016rdj,Donos:2012js,Donos:2013eha,Donos:2014uba,Ling:2015ghh,Ling:2015dma,Ling:2015epa,Ling:2015exa,Ling:2016wyr,Ling:2016dck,Baggioli:2014roa,Baggioli:2016oqk,Baggioli:2016oju,Donos:2014oha,Kiritsis:2015oxa}, people often adopt the so-called operational definition to identify the metallic phase and insulating phase, i.e.,
\begin{itemize}
    \item Metallic phase: $\partial_T\sigma_{DC}<0$.
    \item Insulating phase: $\partial_T\sigma_{DC}>0$.
    \item Critical point (line): $\partial_T\sigma_{DC}=0$.
\end{itemize}

Here we also study the temperature behaviors of DC conductivity, which are shown in the rest plots in Fig.~\ref{DCcalculate}. When the higher-derivative term is absent, i.e., $\alpha_1=0$, $\sigma_{DC}$ is independent of the temperature. Once the coupling parameter $\alpha_1$ is turned on, DC conductivity decreases with $T$ decreasing. In terms of the operational definition of phase of the holographic system described above, this holographic system exhibits the insulating behavior. But we note that unless $\alpha_1$ takes the bounded value $\alpha_1=2/3$ and $\alpha$ tends to infinity, DC conductivity cannot vanish in the limit of zero temperature. For the fixed $\alpha$, $\sigma_{DC}$ decreases with $\alpha_1$ increasing.

In next section, we shall explore the properties of the AC conductivity of this gauge-axion coupling model.

\subsection{AC conductivity}\label{sec-AC}
In this subsection, we numerically calculate AC conductivity in the normal state. Since our system is homogeneous and isotropy, it is enough to turn on the following consistent linear perturbation
\fa
&&
\delta A_x(t,u,x^i)=\int^{+\infty}_{-\infty}\frac{d\omega}{(2\pi)^3}e^{-i \omega t}\delta b_x(u),\nonumber\\
&&
\delta g_{tx}(t,u,x^i)=\int ^{+\infty}_{-\infty}\frac{d\omega }{(2\pi)^3}e^{-i \omega t}u^{-2}h_{tx}(u),\nonumber\\
&&
\delta X^x(t,u,x^i)=\int ^{+\infty}_{-\infty}\frac{d\omega}{(2\pi)^3}e^{-i \omega t}\delta\phi^x (u).
\ffa
Thus we shall have three ordinary differential equations for $h_{tx}, \delta b_x, \delta\phi^x$.
It turns out that near the UV boundary $(u\rightarrow0)$, the asymptotic behavior of the Maxwell field falls in the following form
\fa
&&
\delta b_x = b^{(0)}+b^{(1)}u+\cdot\cdot\cdot.
\ffa
According to holographic dictionary, we read off the conductivity as
\fa
\sigma(\omega)=-\frac{ib^{(1)}}{\omega b^{(0)}}.
\label{sigma}
\ffa
Then, we shall impose the ingoing boundary condition at the horizon to solve the perturbative system and read off the conductivity as the function of the frequency in terms of Eq.~\eqref{sigma}.
\begin{figure}
    \center{
        \includegraphics[scale=0.6]{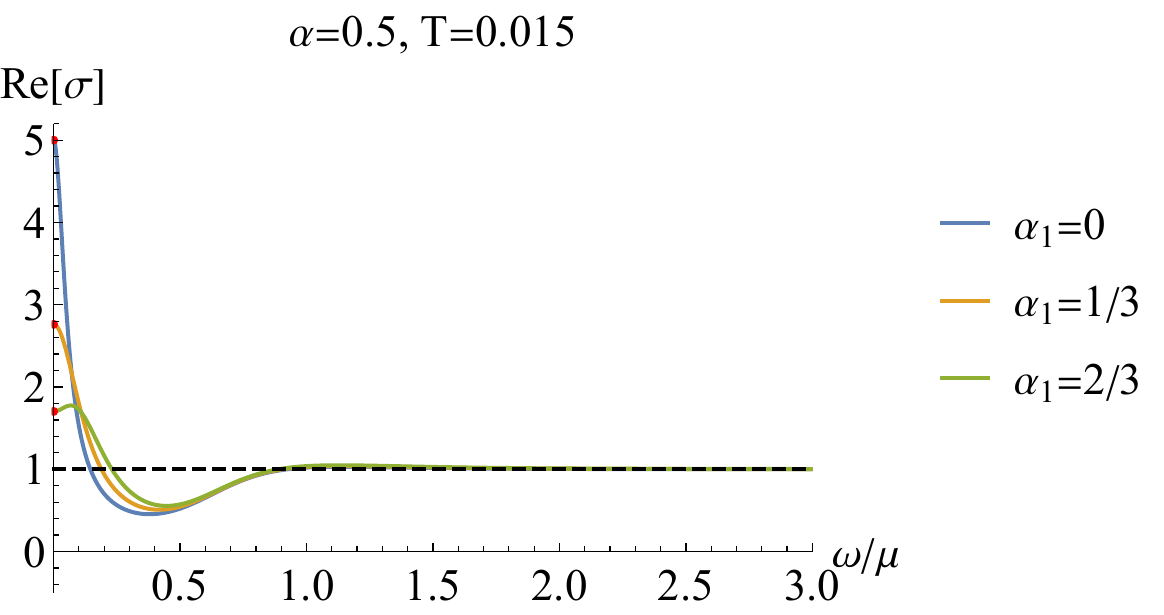}\ \hspace{0.8cm}
        \includegraphics[scale=0.6]{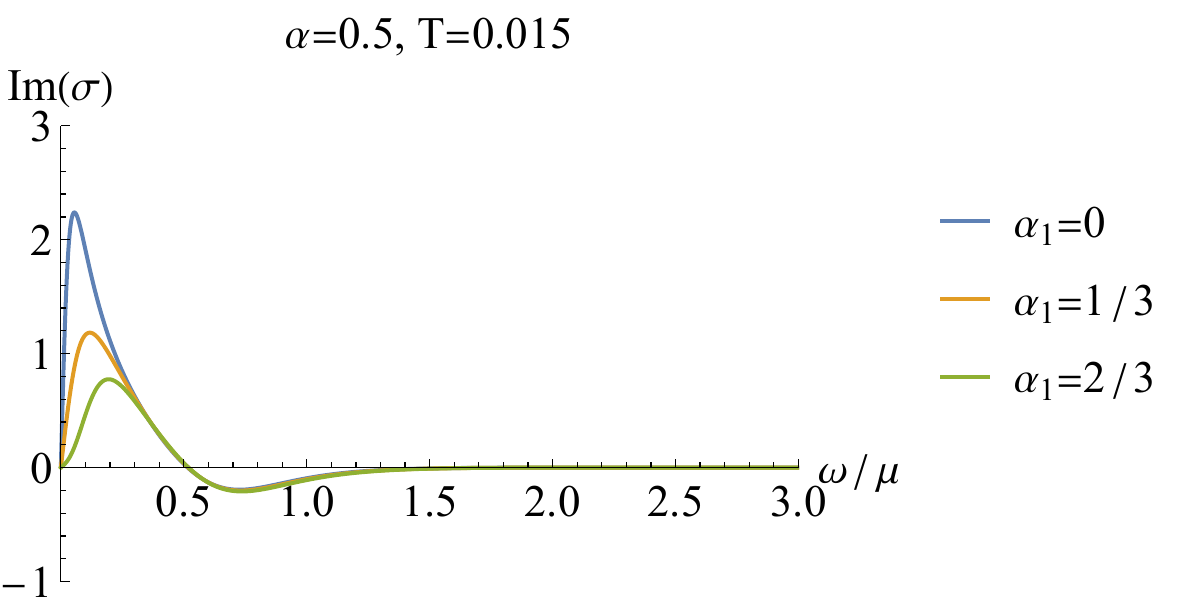}\ \\
        \includegraphics[scale=0.6]{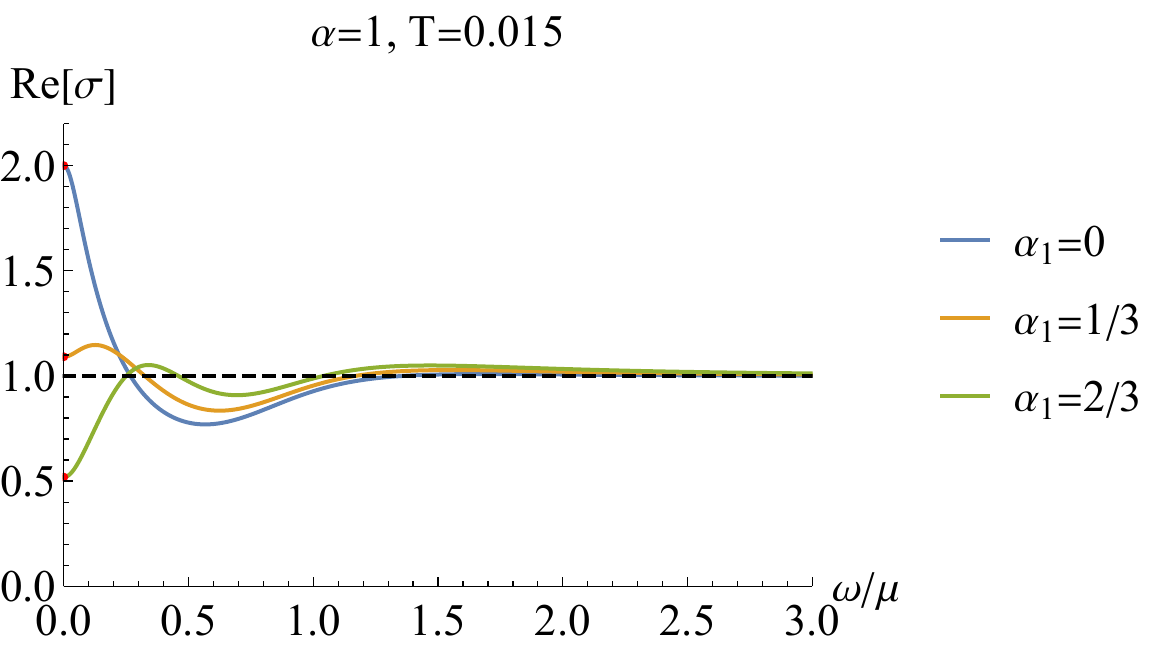}\ \hspace{0.8cm}
        \includegraphics[scale=0.6]{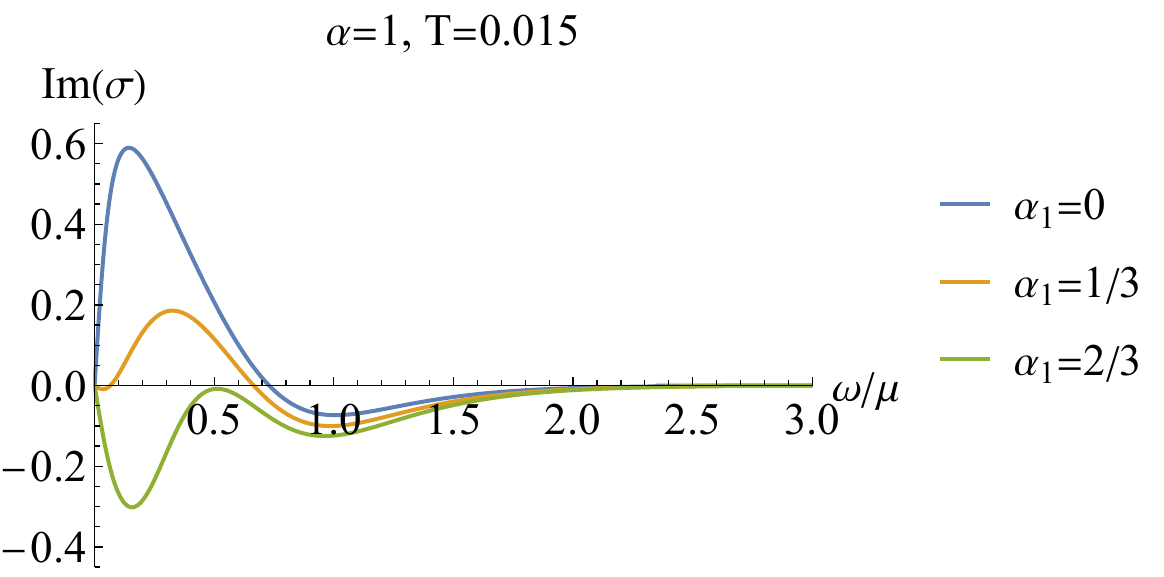}\ \\
        \includegraphics[scale=0.6]{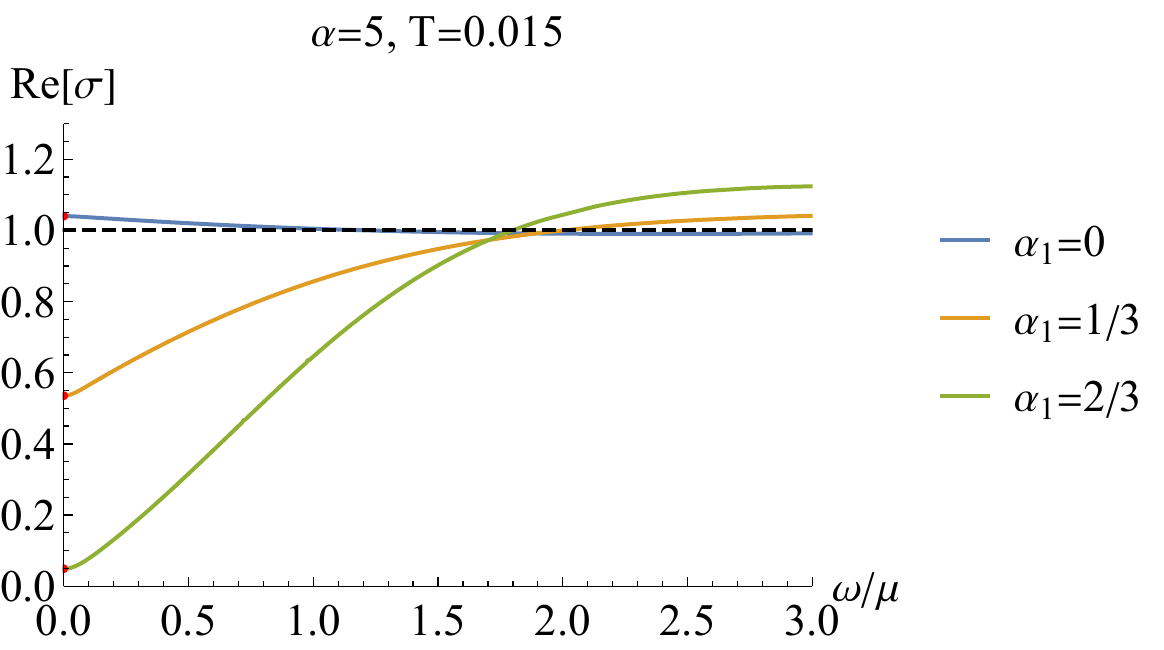}\ \hspace{0.8cm}
        \includegraphics[scale=0.6]{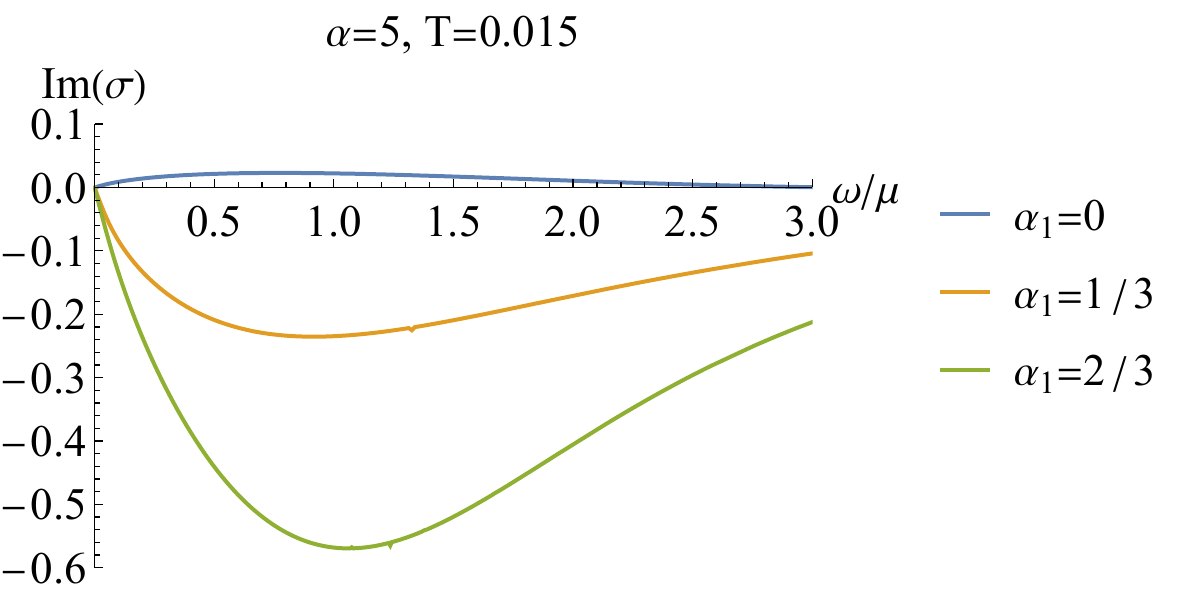}\ \\
        \caption{\label{fig-AC-T0p005} AC conductivity in the normal state at $T= 0.015$ with different $\alpha$ and $\alpha_{1}$. The red dots correspond to DC conductivity obtained from formula (\ref{dc-exp}). }}
\end{figure}

Figure~\ref{fig-AC-T0p005} exhibits the real and imaginary parts of AC conductivity in the normal state at $T= 0.015$. At the high frequency regime, $\omega\gg\mu$, the conductivity tends towards the constant determined by ultraviolet (UV) CFT fixed point. The interesting physics lies at the low frequency region.
Before proceeding, we present some comments on the AC conductivity behavior at low frequency without gauge-axion coupling \cite{Andrade:2013gsa}, i.e., $\alpha_{1}=0$. For this case, we observe that AC conductivity in real part displays a standard Drude peak at low frequency when the momentum dissipation is weak, i.e., $\alpha$ is small. It implies that the weak momentum dissipation drives the current into a coherent state. Theoretically, it is because the total momentum of the system is approximately conserved, for which we can implement a perturbative expansion in the small momentum relaxation rate within the memory matrix formalism \cite{Hartnoll:2007ih,Hartnoll:2012rj,Lucas:2015pxa,Mahajan:2013cja}. With the strength of the momentum dissipation $\alpha$ increasing, the Drude peak gradually reduces and we observe a transition from coherent phase to incoherent phase (see the blue curves in Fig.~\ref{fig-AC-T0p005}, also see Refs.~\cite{Kim:2014bza,Wu:2018zdc}). In fact, because we must include the contribution at subleading order in the relaxation rate when the momentum dissipation is strong, the behavior of AC conductivity at low frequency should be depicted by a modified holographic formula, first derived in \cite{Davison:2015bea} and generalized to a more general holographic theory \cite{Zhou:2015qui}, instead of the standard Drude formula or even the modified hydrodynamic results. However, when the gauge-axion coupling term $\alpha_1 Tr[XF^{2}]$ in the Lagrangian density \eqref{lagarangex} is introduced, AC conductivity at low frequency exhibits some interesting behaviors.

We first discuss the case of weak momentum dissipation ($\alpha=0.5$), for which with $\alpha_1$ increasing, the peak at low frequency gradually deviates from the Drude form (see the orange curves in the first row in Fig.~\ref{fig-AC-T0p005}). Recalling that in the holographic higher derivative theory of a probe Maxwell field coupled to the Weyl tensor $C_{\mu\nu\rho\sigma}$ over the Schwarzschild-AdS (SS-AdS) black brane background \cite{Myers:2010pk,Sachdev:2011wg,Hartnoll:2016apf,Ritz:2008kh,Witczak-Krempa:2012qgh,Witczak-Krempa:2013xlz,Witczak-Krempa:2013nua,Witczak-Krempa:2013aea,Katz:2014rla,Wu:2018xjy}, depending on the sign of the coupling parameter, AC conductivity at low frequency exhibits a so-called Damle-Sachdev (DS) peak being similar to the particle response or a dip being similar to the vortex response \cite{Damle:1997rxu}. A similar DS peak has also been observed in probe branes and DBI action \cite{Chen:2017dsy} and just higher terms in $F^2$ \cite{Baggioli:2016oju}. Similar to the case of Weyl theory studied in Refs.~\cite{Myers:2010pk,Sachdev:2011wg,Hartnoll:2016apf,Ritz:2008kh,Witczak-Krempa:2012qgh,Witczak-Krempa:2013xlz,Witczak-Krempa:2013nua,Witczak-Krempa:2013aea,Katz:2014rla,Wu:2018xjy}, the gauge-axion coupling term with $\alpha_1>0$ induces a behavior similar to the vortex response, which exhibits a dip at low frequency in AC conductivity. Therefore, we conclude that in the weak momentum dissipation, the gauge-axion coupling term drives the behavior of AC conductivity at low frequency away from the standard Drude peak. That is to say, there is a competition between the momentum dissipation and the gauge-axion coupling term, and finally the effect of momentum dissipation dominates over that of the gauge-axion coupling.

Further increasing $\alpha_1$ to $\alpha_1=2/3$, a small but evident pronounced peak emerges at intermediate frequency (see the green curves in the first row in Fig.~\ref{fig-AC-T0p005}). Such pronounced peak has been observed in our previous work \cite{Wu:2018xjy}, for which we attribute the pronounced peak to the coupling among the nontrivial scalar field dual to a relevant operator, Maxwell field and the Weyl tensor. Though at this moment, we cannot analytically work out the expression of the pronounced peak such that we cannot clearly know the mechanism of this pronounced peak. However it can be still believed that the pronounced peak observed here shares the similar origin with that in Ref.~\cite{Wu:2018xjy}. Therefore, we argue that the pronounced peak studied here is associated with the gauge-axion coupling. In future, we expect that the origin of the pronounced peak at intermediate frequency can be clearly illustrated.

In addition, we find that for $\alpha=0.5$ and $\alpha_1=2/3$, an obvious spectral weight transfer from the low frequency to the intermediate frequency. A similar spectral weight transfer is also observed in the holographic higher-derivative theory studied in Ref.~\cite{Wu:2018xjy}. In particular, the phenomenon of the spectral weight transfer occurs only in the holographic  theory with the response similar to the vortex behavior and does not happen in that with response similar to the particle response \cite{Wu:2018xjy}.

Then we increase the strength of momentum dissipation to $\alpha=1$. The pronounced peak and the spectral weight transfer have appeared for $\alpha_1=1/3$. With further increasing $\alpha_1$ to $\alpha_1=2/3$, a dip emerges. It is because with the increase of the strength of momentum dissipation, there is a transition from coherent phase to incoherent one leading to the emergence of a dip instead of the Drude-like peak.

There is no doubt that with the strength of momentum dissipation further increasing (see the last row in  Fig.~\ref{fig-AC-T0p005} for $\alpha=5$), a dip begin to emerge only small $\alpha_1$ is turned on. With $\alpha_1$ increasing, a dip becomes deeper. Therefore, we conclude that the dip is the result of a combination of the strong momentum dissipation and the gauge-axion coupling.

\section{Superconducting phase}\label{The Superconducting Phase}

In this section, we turn to study the superconducting properties of this holographic effective theory with gauge-axion coupling. Some works have studied the superconducting properties of holographic axion model \cite{Andrade:2014xca,Kim:2015dna,Baggioli:2015dwa,Kiritsis:2015hoa,Cai:2020nyd}. Here, we shall mainly focus on the effect of gauge-axion coupling.

\subsection{Condensation}

Under the ansatz \eqref{bhbs}, we numerically solve the equations of motion (\ref{maxwell-eq}) -- (\ref{einstein-eq}) to explore the properties of the condensation. Again, it is easy to find that the gauge-axion coupling term $J {\rm Tr}[XF^{2}]$ has no any effect on the background solution even in the superconducting phase. Only three parameters: the strength of momentum dissipation $\alpha$, the gauge coupling parameters $a$ and $n$,
exert their influence on the condensation. Notice that from the Lagrangian density $\mathcal{L}_{\chi}$, i.e., Eq.~\eqref{Lchi}, we see that the relevant quantity is the product of $q$ and $n$. At the same time, since the increase of charge make the condensation easier, which had been widely studied in previous references for example \cite{Horowitz:2013jaa,Ling:2014laa}, we shall set $q=2$ and $n=1$ through this paper without loss of generality.

\begin{figure}
    \center{
        \includegraphics[scale=0.43]{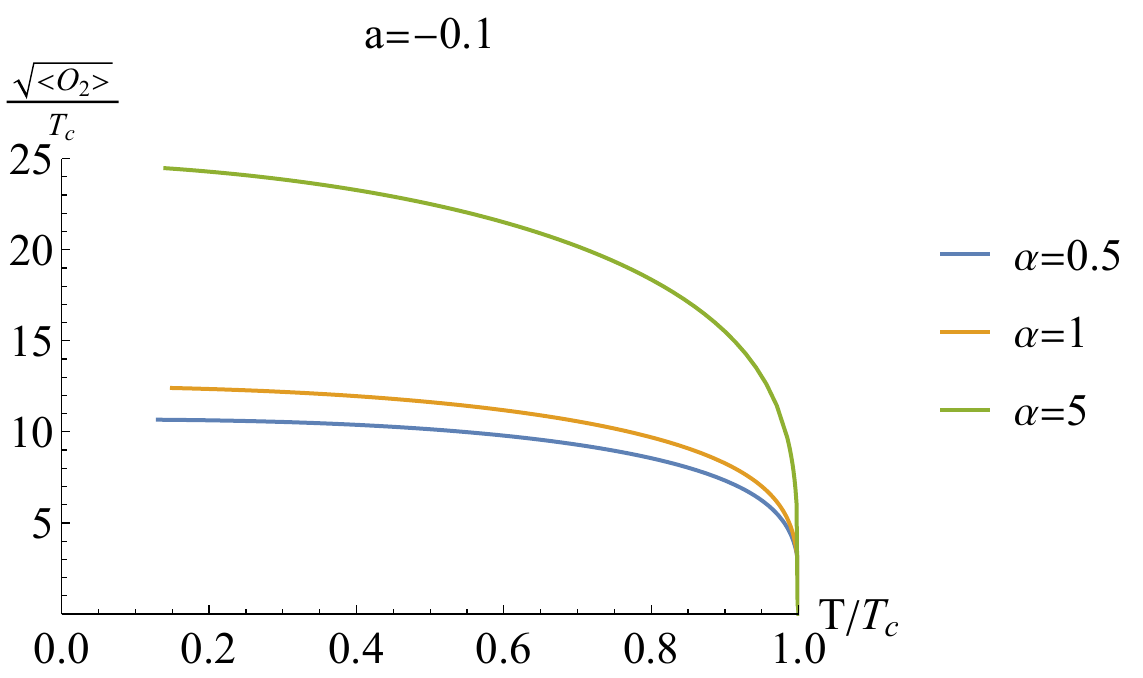}\hspace{0.1cm}
        \includegraphics[scale=0.43]{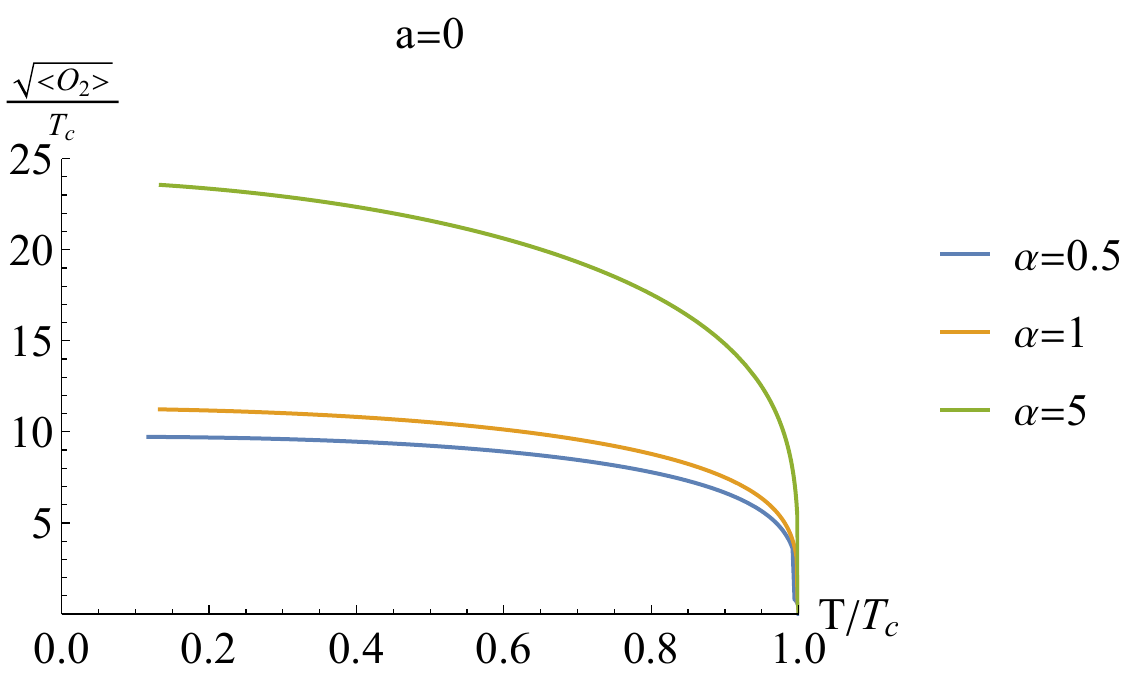}\hspace{0.1cm}
        \includegraphics[scale=0.43]{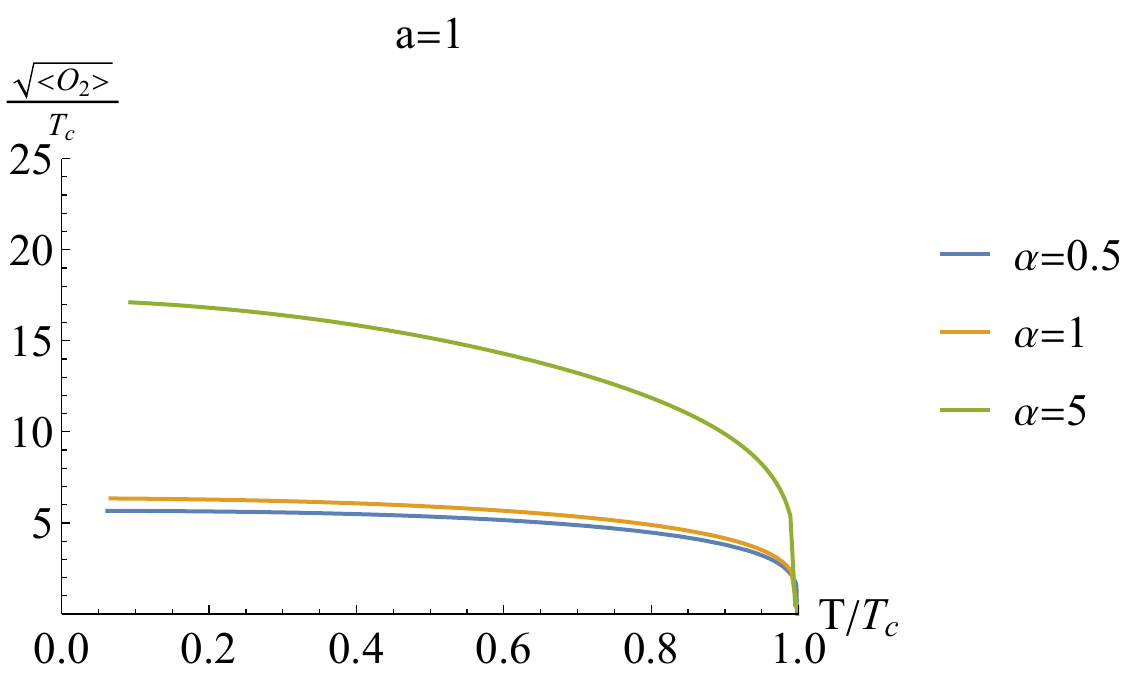}\ \\
        \caption{\label{fig-O2} The condensation $<O_2>$ as a function of temperature for given $a$ (from left to right $a=-0.1$, $a=0$ and $a=1$) and various values of $\alpha$. We have set $q=2$ and $n=1$.}}
\end{figure}

In the unit of the critical temperature we plot the condensation $<O_2>$ as a function of the temperature by fixing the gauge coupling parameter $a$ or by fixing the strength of momentum dissipation $\alpha$, respectively. When we fix the gauge coupling parameter $a$, the stronger the strength of momentum dissipation, the expectation value of the condensation becomes much larger and thus the critical temperature becomes lower (Fig. \ref{fig-O2}), which had also been observed in many works for example \cite{Kim:2015dna,Zeng:2014uoa,Ling:2014laa}. Therefore, we conclude that the condensation becomes harder with the strength of momentum dissipation increasing even in presence of the gauge coupling. We expect that the condensation is violated as the momentum dissipation becomes stronger beyond some critical values.

\begin{figure}
    \center{
        \includegraphics[scale=0.43]{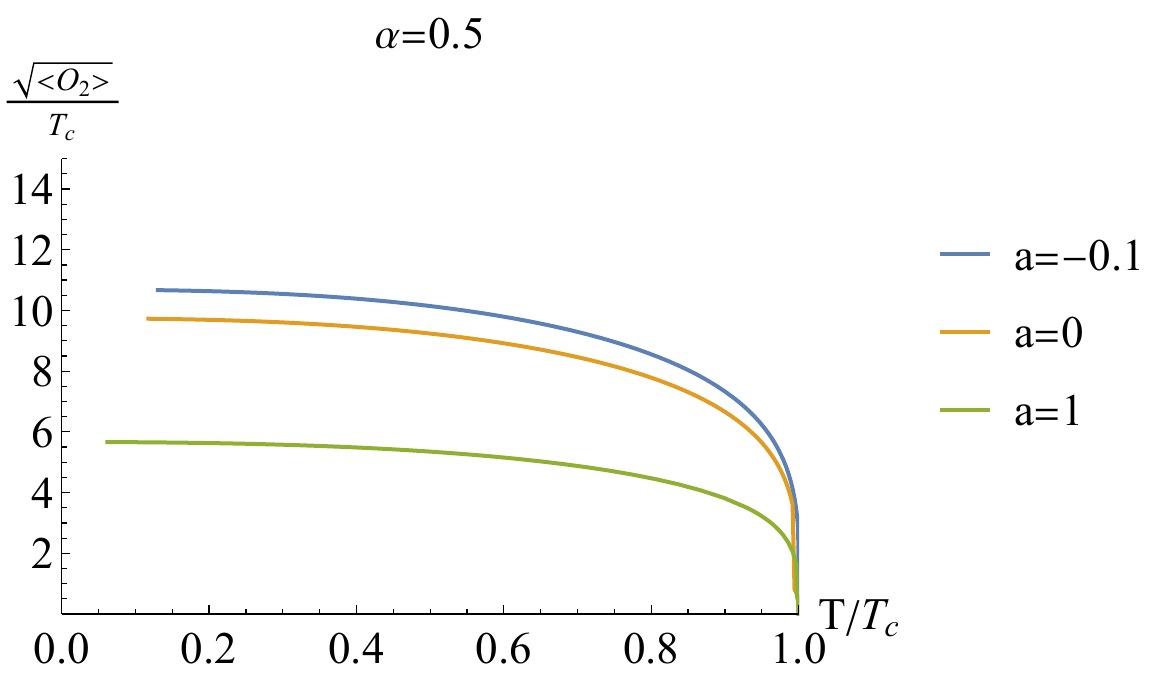}\hspace{0.1cm}
        \includegraphics[scale=0.43]{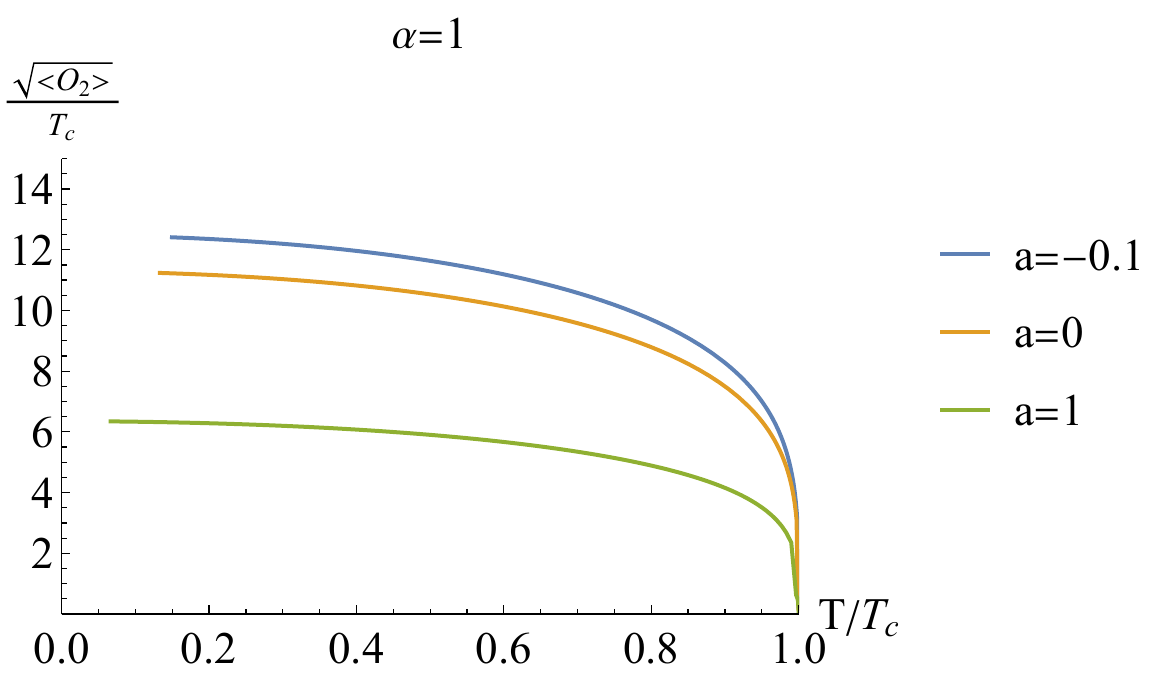}\hspace{0.1cm}
        \includegraphics[scale=0.43]{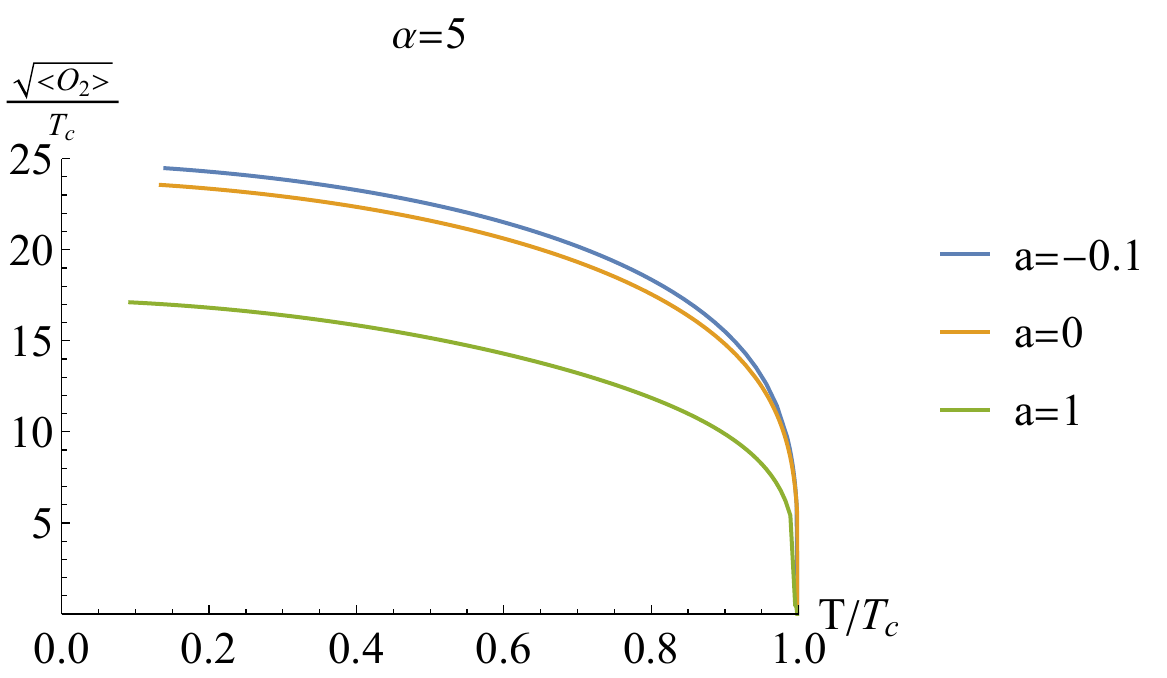}\
        \caption{\label{fig-O2-v2} The condensation $<O_2>$ as a function of temperature for given $\alpha$ (from left to right $\alpha=0.5$, $\alpha=1$ and $\alpha=5$) and various values of $a$. We have set $q=2$ and $n=1$.}}
\end{figure}

And then, we fix the strength of momentum dissipation $\alpha$ and vary the gauge coupling parameter $a$ to see the effect of gauge coupling on the condensation. From Fig. \ref{fig-O2-v2}, we see that with $a$ increasing, the expectation value of the condensate becomes much smaller. It indicates that the condensation becomes easier and the critical temperature becomes higher.

\subsection{Superconducting conductivity}

To calculate AC conductivity in the superconducting phase, we can follow the same procedure in the normal state outlined in Sec.~\ref{sec-AC}.
Different from the condensation, there are four theory parameters ($\alpha$, $a$, $\alpha_{1}$, $\alpha_{2}$) affecting the superconducting conductivity.
In what follows, we shall explore the effects of these coupling parameters on the conductivity step by step.

\begin{figure}
    \center{
        \includegraphics[scale=0.6]{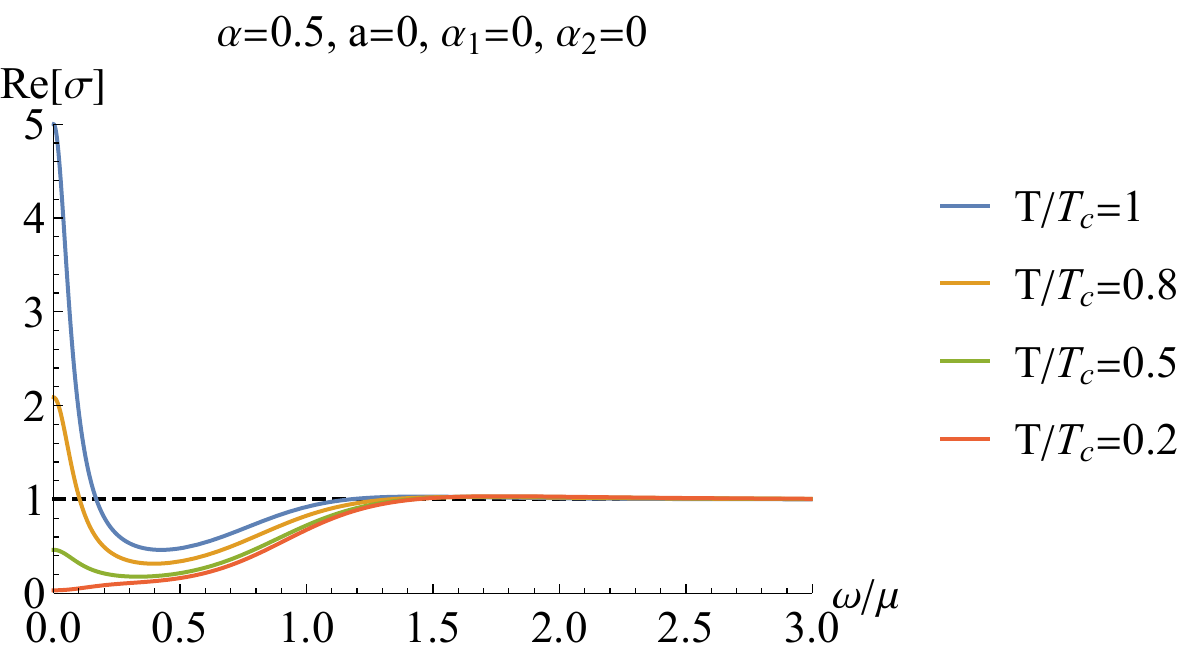}\ \hspace{0.8cm}
        \includegraphics[scale=0.6]{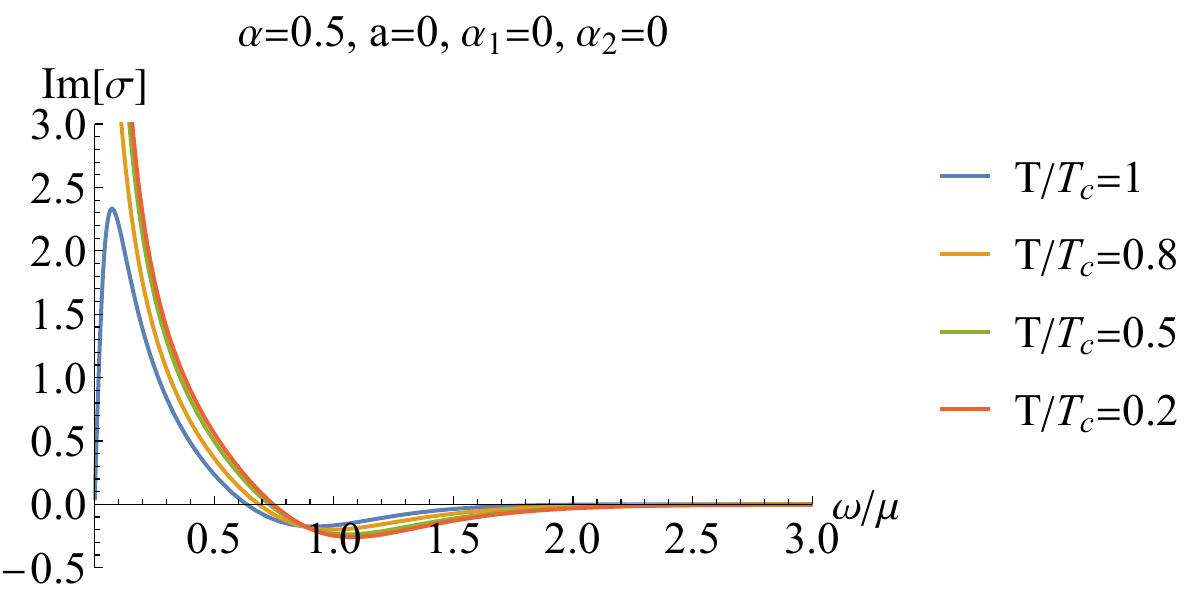}\ \\
        \includegraphics[scale=0.6]{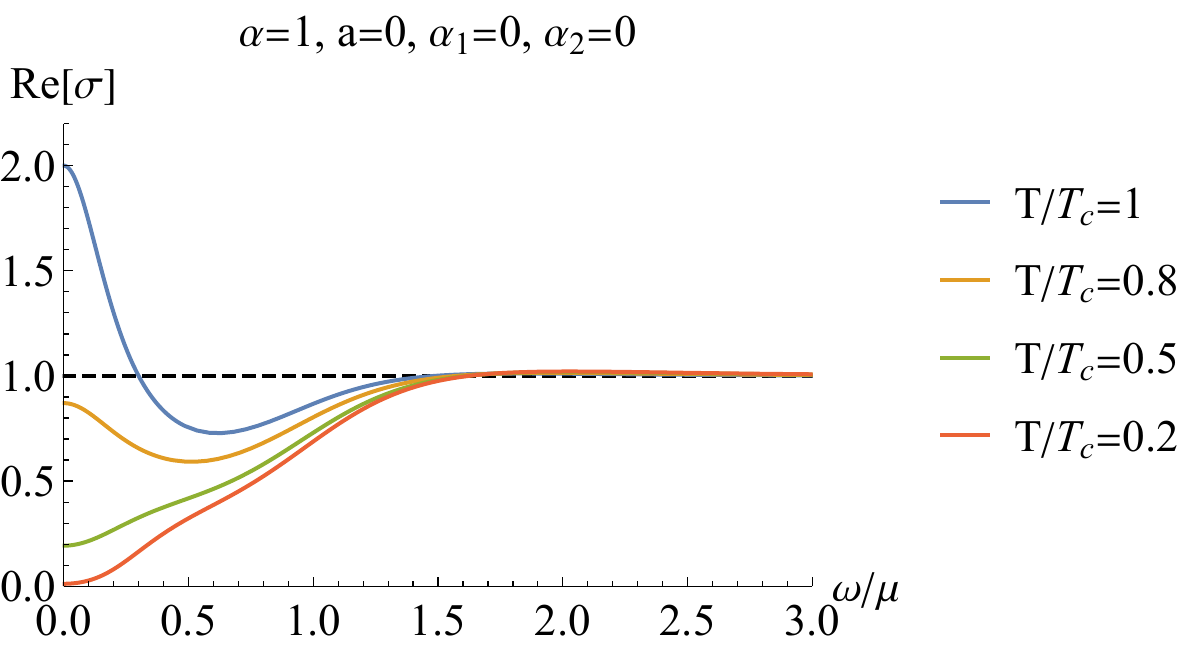}\ \hspace{0.8cm}
        \includegraphics[scale=0.6]{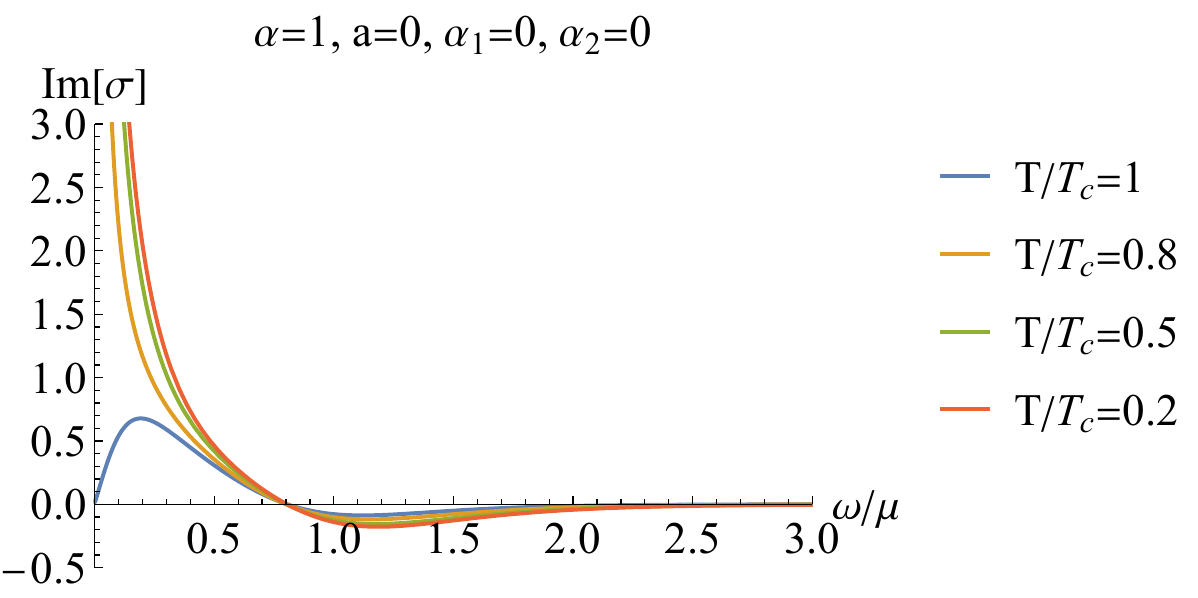}\ \\
        \includegraphics[scale=0.6]{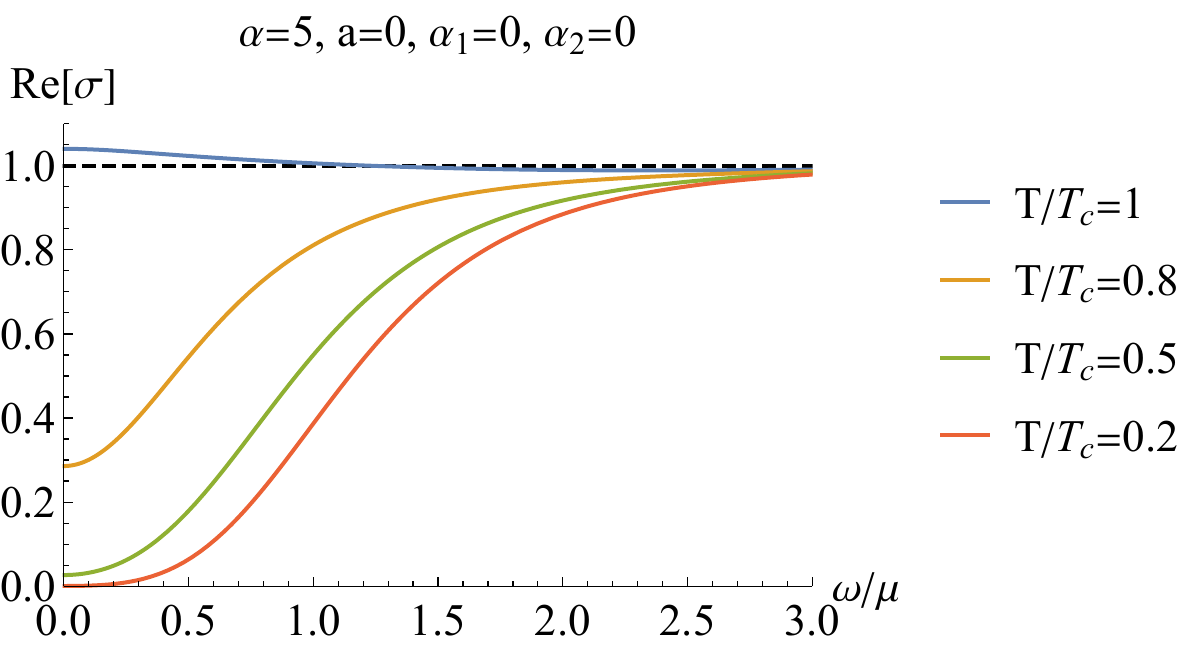}\ \hspace{0.8cm}
        \includegraphics[scale=0.6]{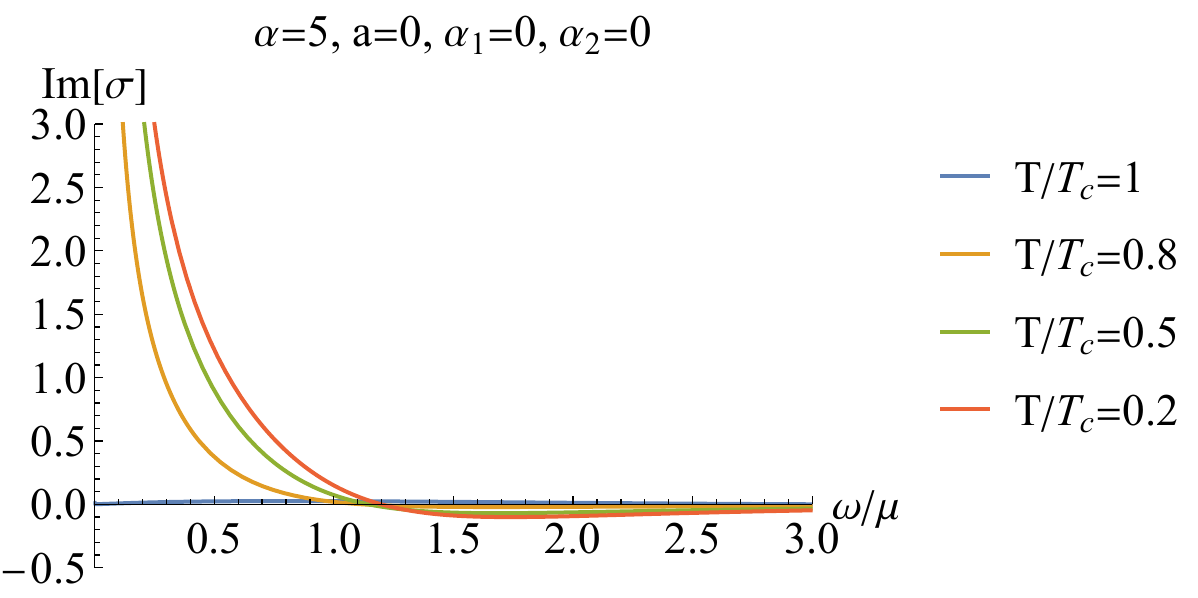}\ \\
        \caption{\label{fig-SC-nalist} The conductivities $\sigma$ as a function of the frequency for different $\alpha$ fixing $a=0$, $\alpha_{1}=0$, and $\alpha_{2}=0$. The left panels are for the real part and the right panels are for the imaginary one. }}
\end{figure}

\begin{figure}
    \center{
        \includegraphics[scale=0.43]{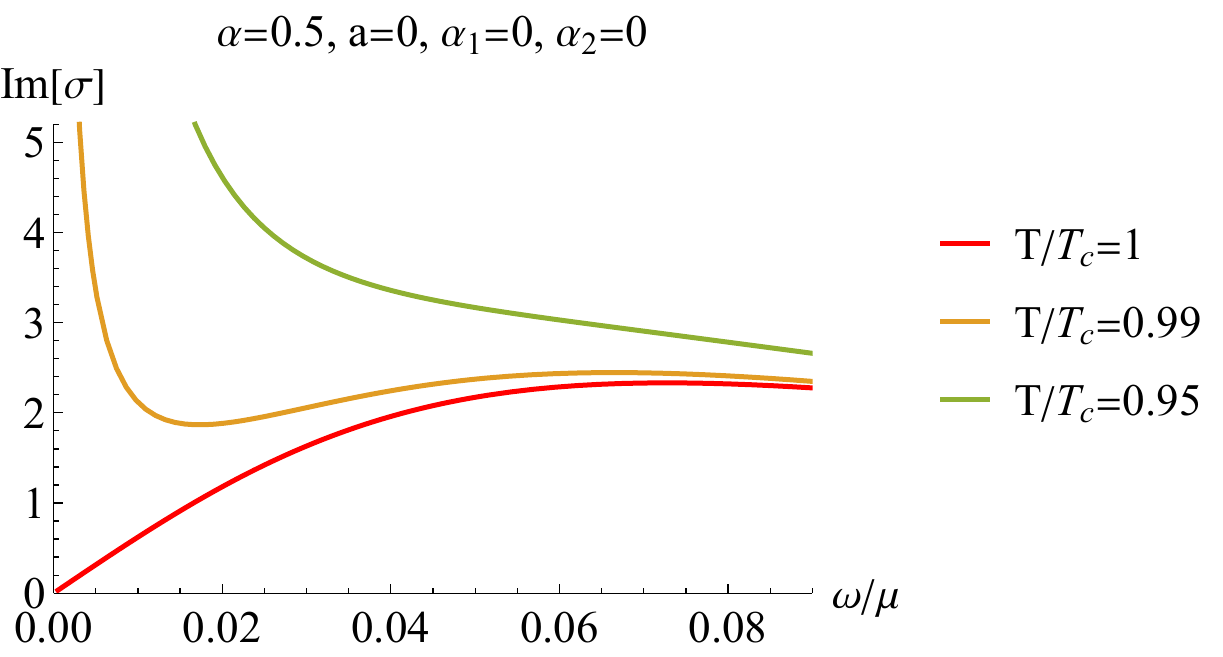}\hspace{0.1cm}
        \includegraphics[scale=0.43]{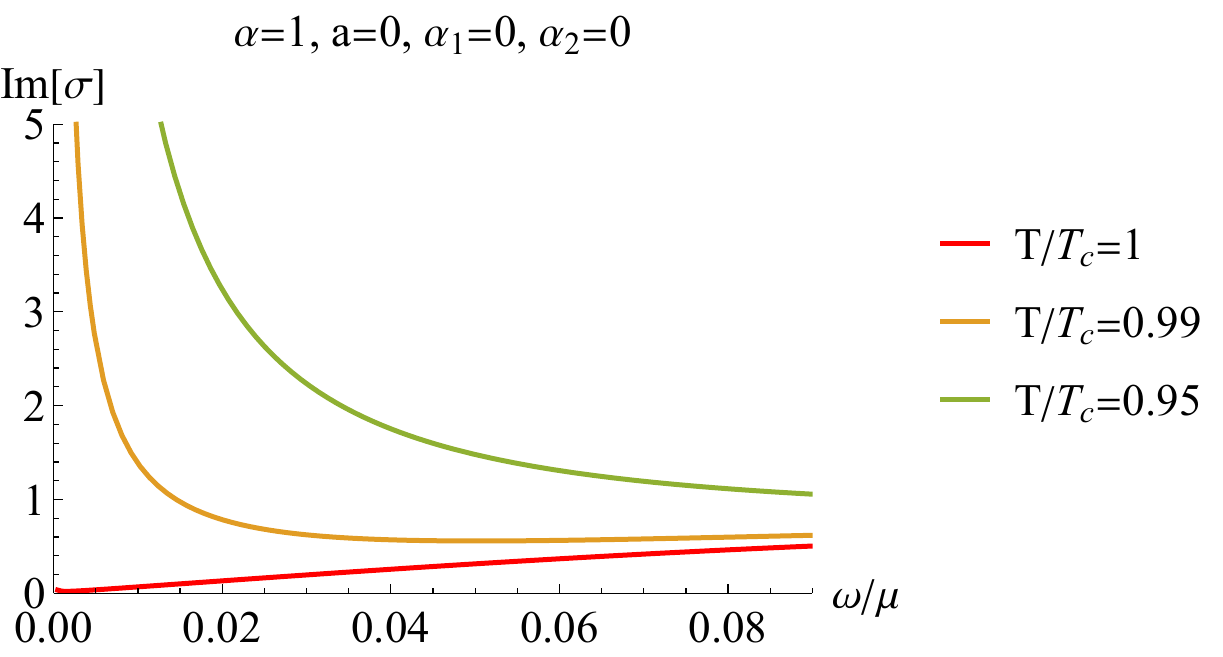}\hspace{0.1cm}
        \includegraphics[scale=0.43]{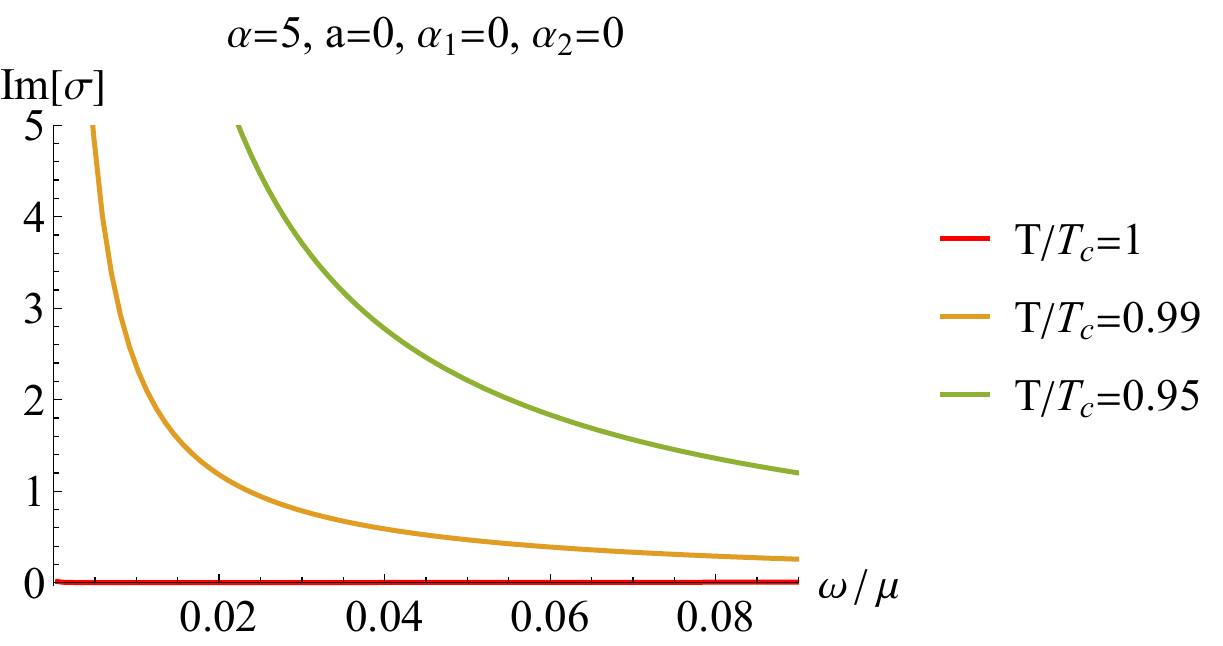}\ \\
        \includegraphics[scale=0.43]{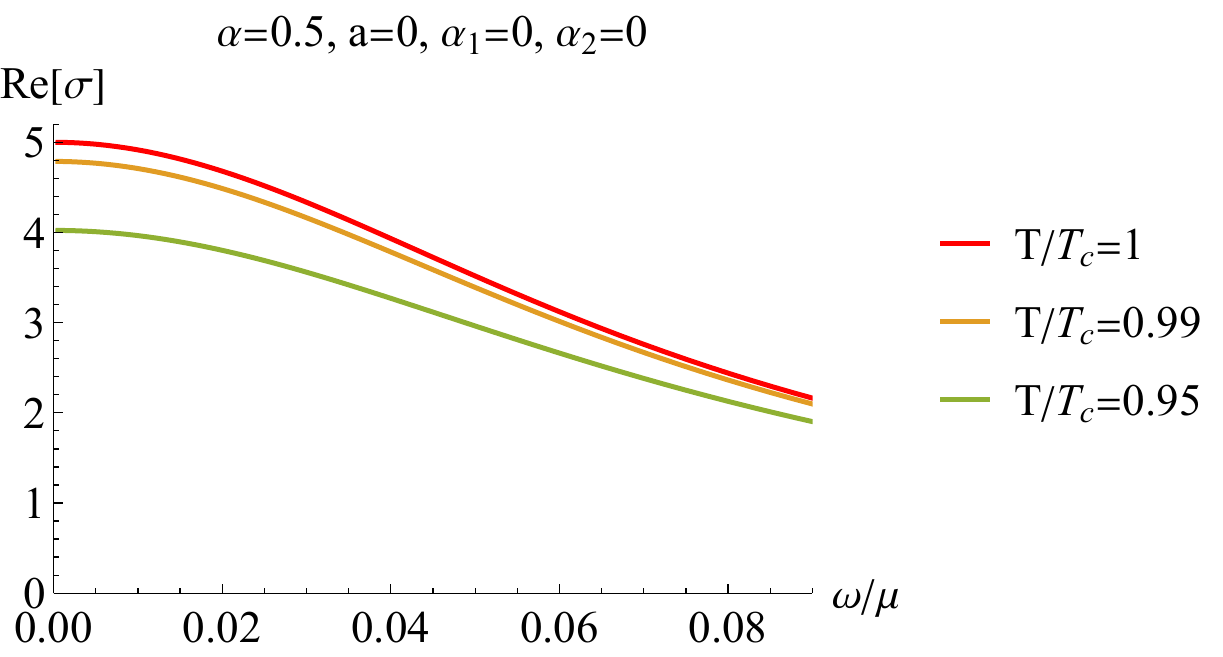}\hspace{0.1cm}
        \includegraphics[scale=0.43]{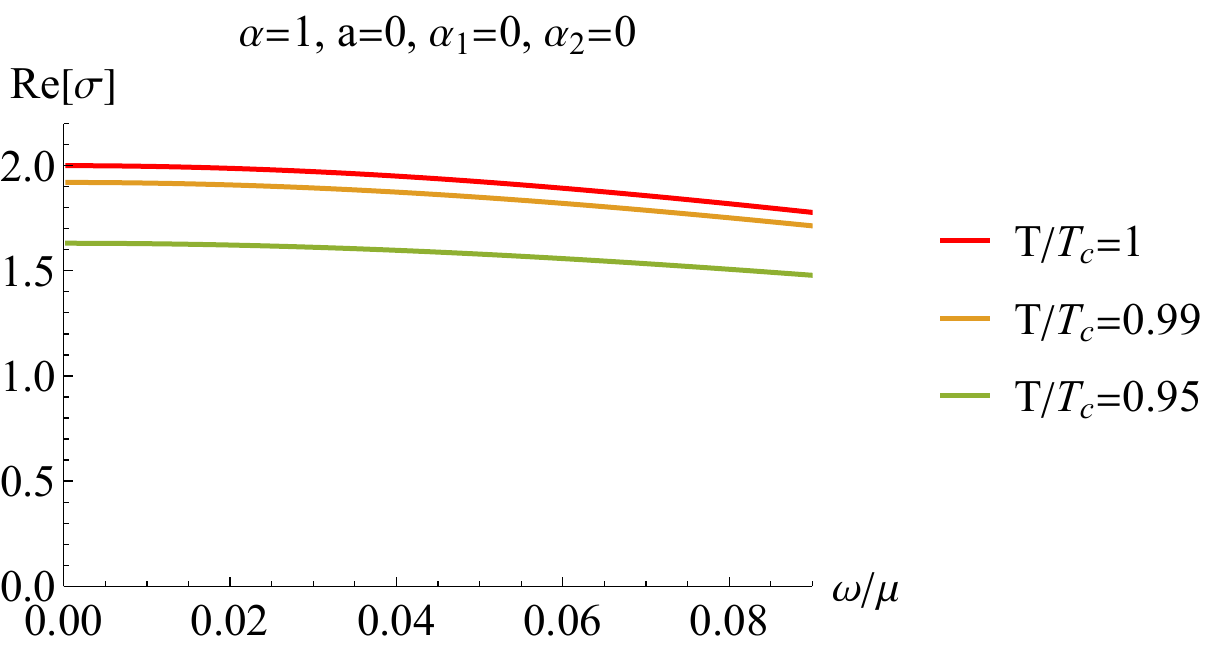}\hspace{0.1cm}
        \includegraphics[scale=0.43]{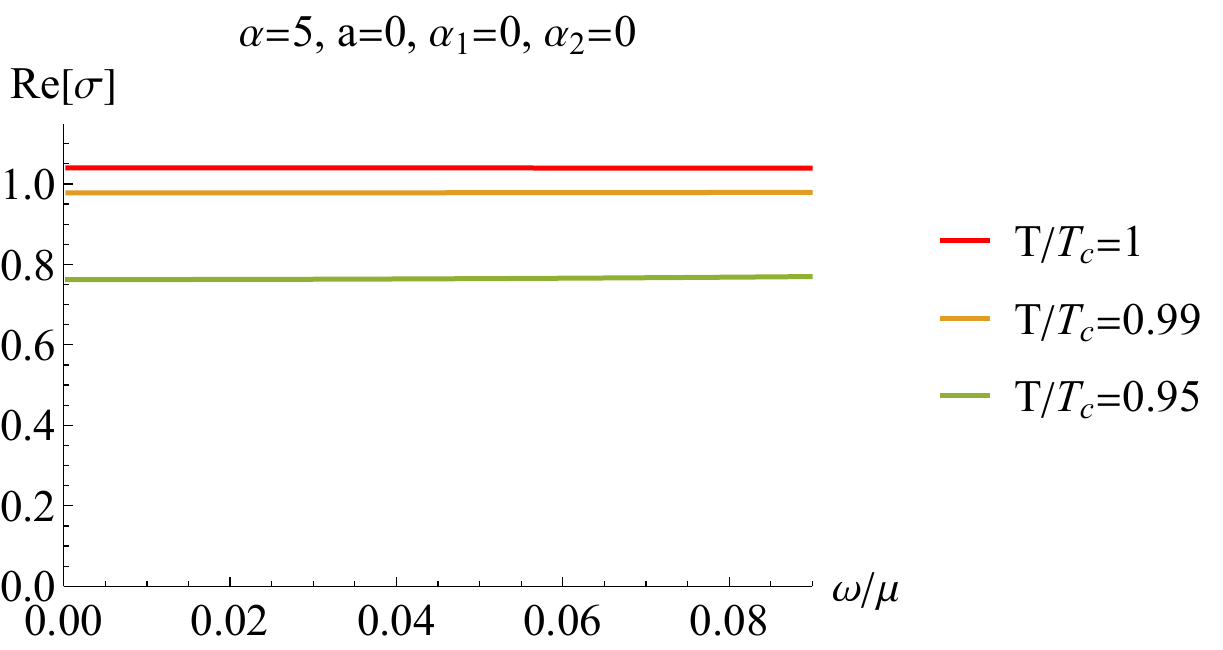}\
        \caption{\label{near Tc} The critical behavior of the conductivity near the critical temperature in the low frequency region for the different $\alpha$. We have turned off other coupling parameters.}}
\end{figure}

We first study the effect of the momentum dissipation on AC conductivity. To this end, we turn off the other coupling parameters. In Fig.~\ref{fig-SC-nalist}, we plot the real and imaginary parts of AC conductivity as a function of frequency for various $\alpha$. We also show the evolution of the conductivity with the temperature from normal phase to superconducting phase. We observe that once the system enters into the superconducting phase, the imaginary parts of the conductivity climb up rapidly and tends to infinity in the limit of $\omega=0$. According to the Kramers-Kronig (KK) relation, a pole emerges in $Im[\sigma]$ implying a corresponding delta function in $Re[\sigma]$, which is one of the hallmarks of genuine superconductor. In order to exhibit such sudden change more distinctly, we plot the conductivity in low frequency region with the temperature dropping through the critical point . Unlike the abrupt changing behavior of the imaginary part, the real part of the conductivity only changes slightly near the critical temperature. The real part of optical conductivity at low frequency goes down with the temperature decreasing and finally vanishes at extremal low temperature. Therefore, our holographic model also resembles a two-fluid model as the standard holographic superconductor model \cite{Hartnoll:2008vx,Hartnoll:2008kx,Horowitz:2010gk}, and the holographic Q-lattice superconductor \cite{Ling:2014laa}, and the holographic superconductor from higher derivative theory \cite{Wu:2017xki,Liu:2020hhx}.
\begin{figure}
    \center{
        \includegraphics[scale=0.42]{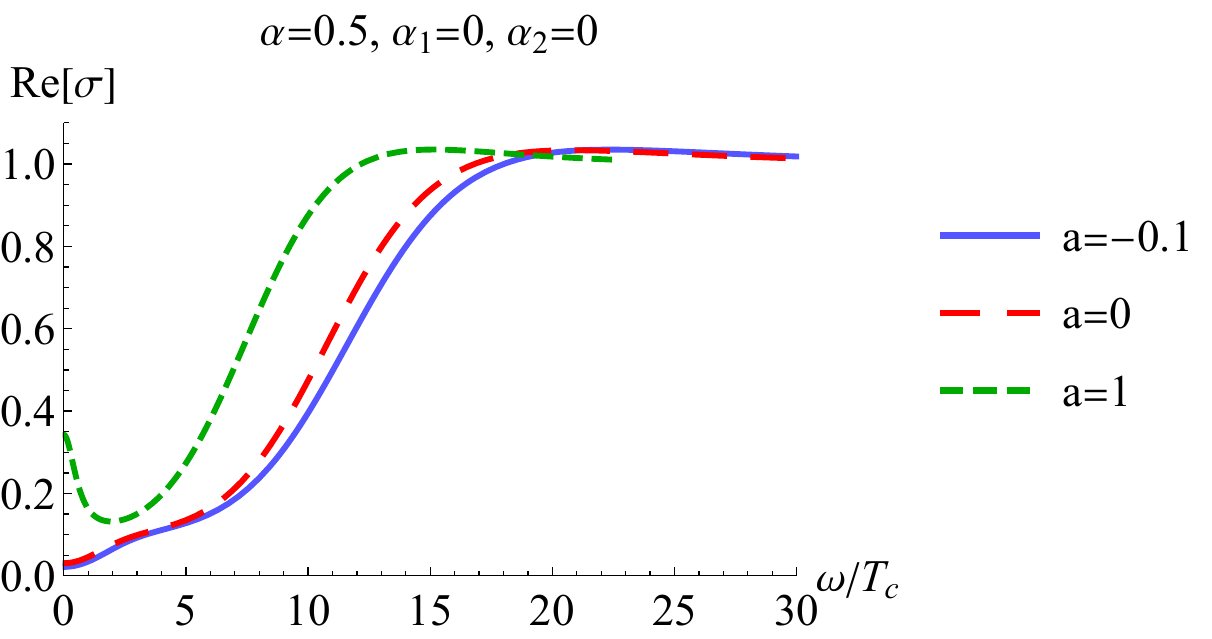}\hspace{0.1cm}
        \includegraphics[scale=0.42]{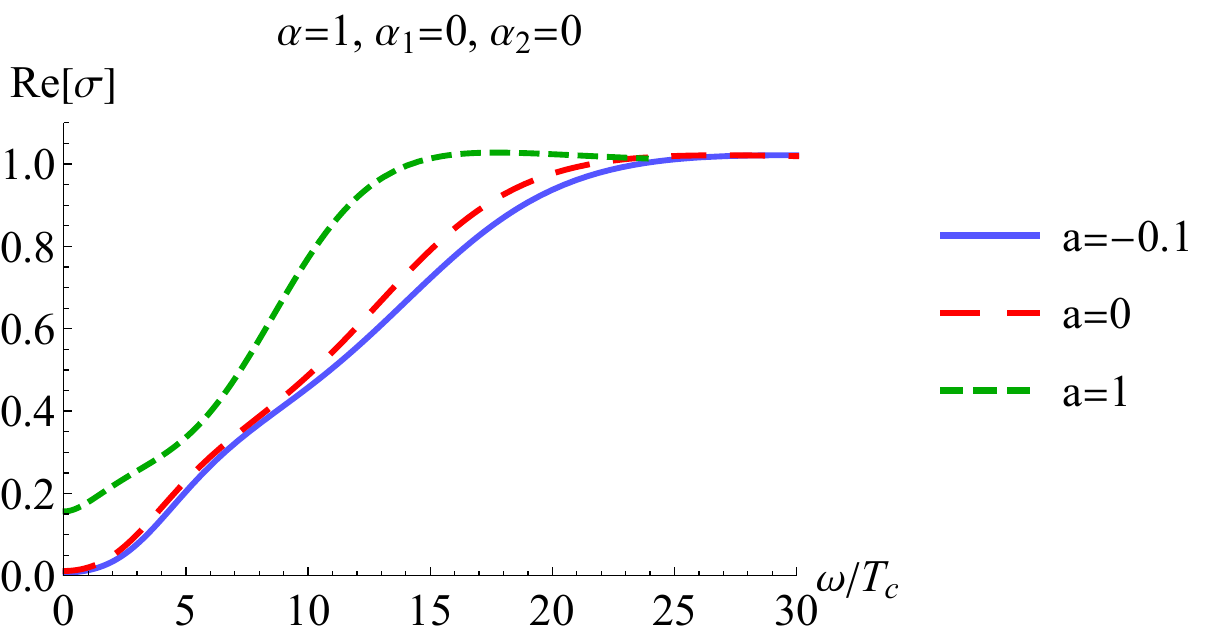}\hspace{0.1cm}
        \includegraphics[scale=0.42]{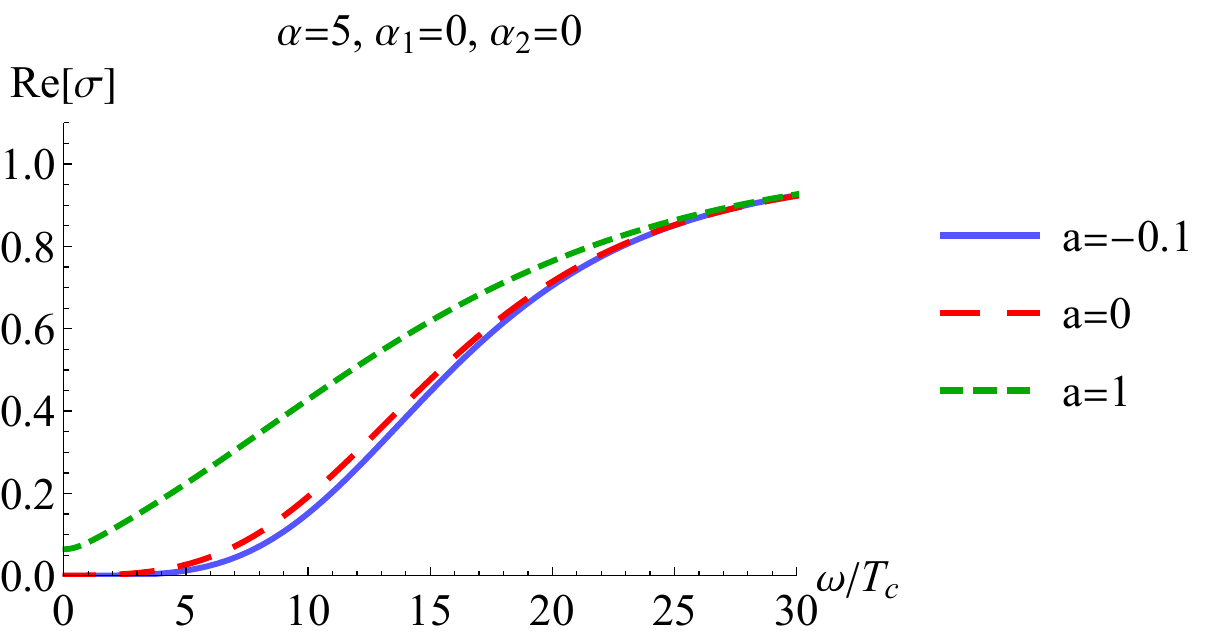}\ \\
        \includegraphics[scale=0.42]{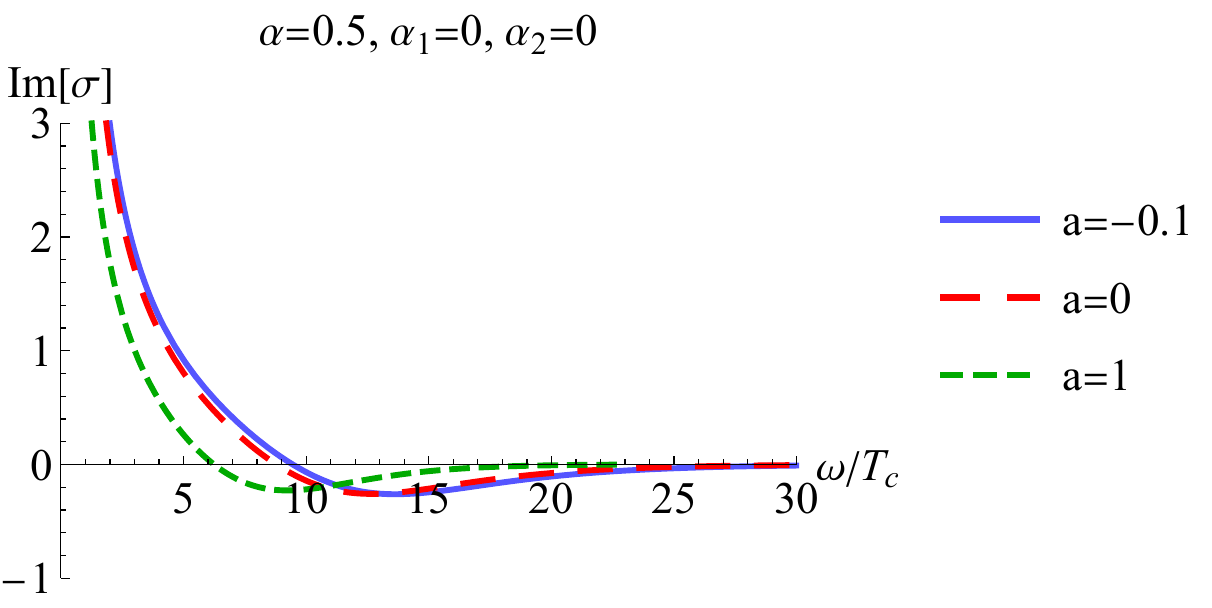}\hspace{0.1cm}
        \includegraphics[scale=0.42]{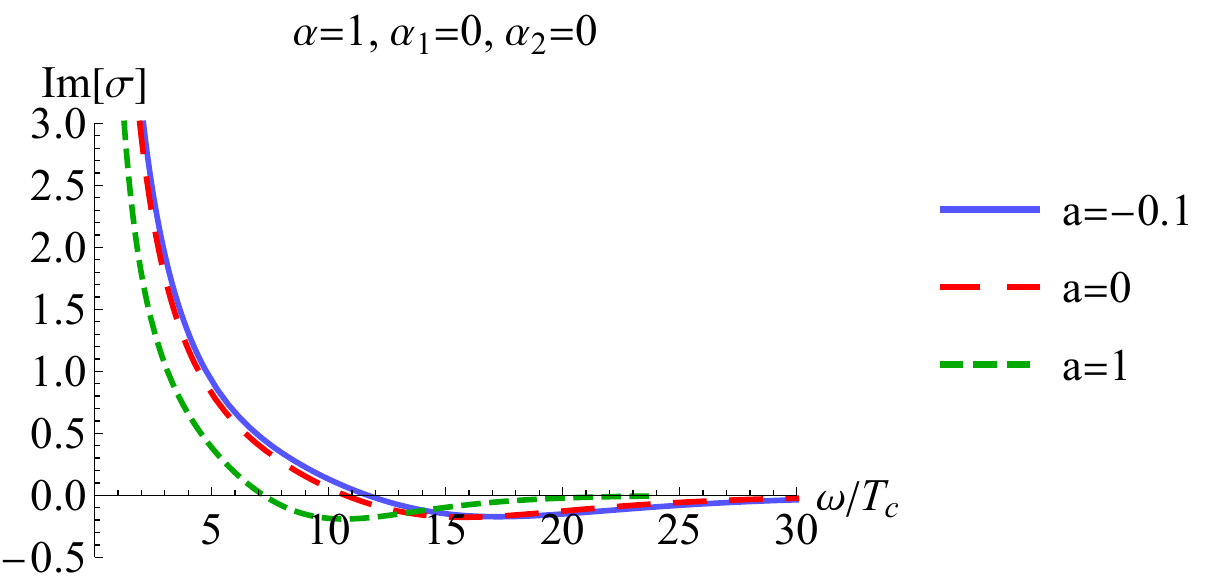}\hspace{0.1cm}
        \includegraphics[scale=0.42]{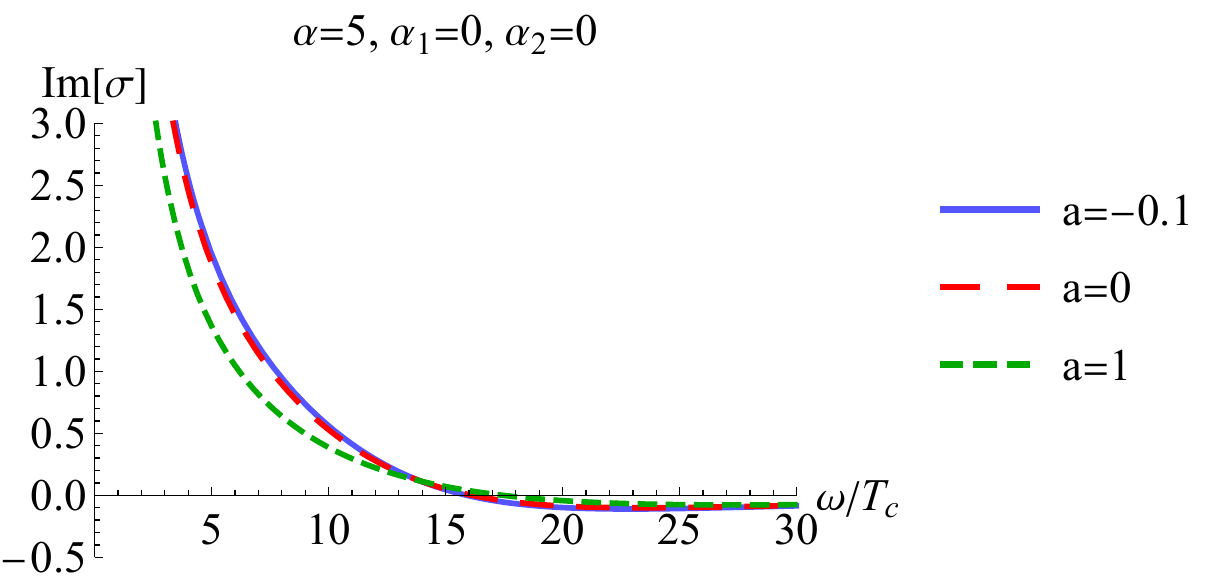}
        \caption{\label{a vs na} The behavior of conductivity with varying $a$ for different $\alpha$ in the superconducting state. Here we fix the temperature $T/T_{c}\approx0.2$. }}
\end{figure}

Furthermore, we study the combined effect of the strength of momentum dissipation $\alpha$ and the gauge coupling parameter $a$ in the superconducting state. Here we have fixed the remaining parameters ($\alpha_{1}$ and $\alpha_{2}$) to be zero. In Fig.~\ref{a vs na}, we show both real and imaginary parts of conductivity with varying $a$ for different $\alpha$. At low temperature ($T/T_{c}\approx0.2$), the real part of the conductivity tends to zero in the limit of $\omega\rightarrow 0$ (see the dashed red lines in Fig.~\ref{a vs na} and also see Fig.~\ref{fig-SC-nalist}). It implies that at this temperature, most of the normal components of the electron fluid have formed the superfluid component. For negative $a$ ($a=-0.1$), the case is similar to that of $a=0$. However, for positive $a$ ($a=1$), the DC conductivity starts to rise up. It means that we need further cool the system to drive the normal component of the electron fluid forming the superfluid component.

\begin{figure}
    \center{
        \includegraphics[scale=0.62]{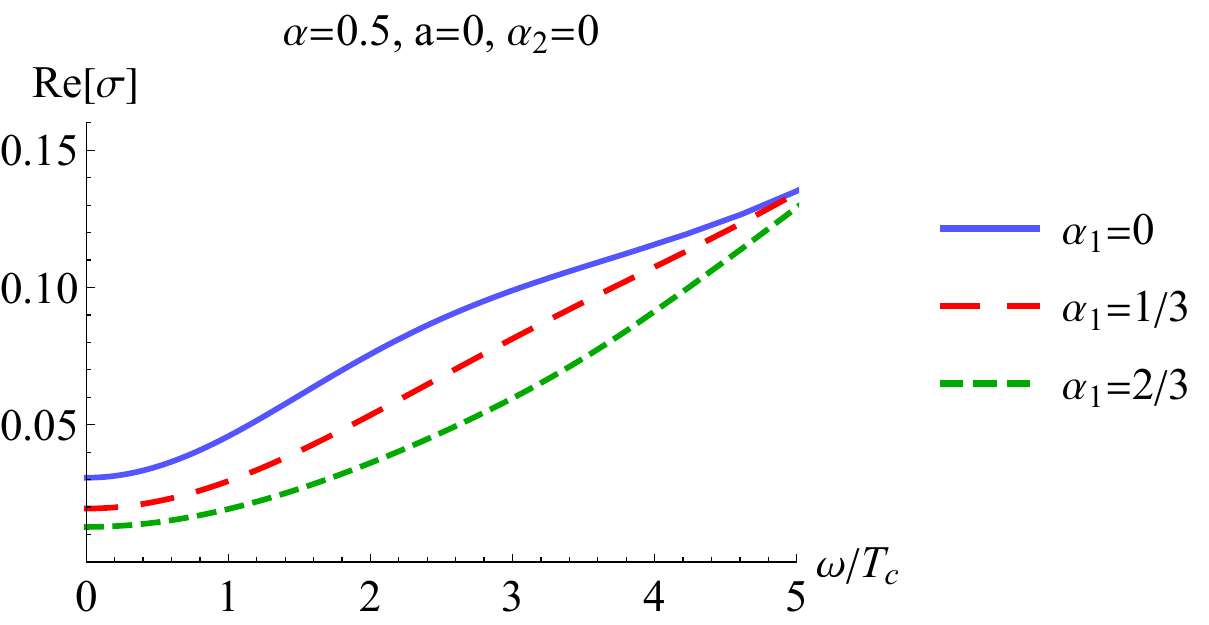}\hspace{0.5cm}
        \includegraphics[scale=0.62]{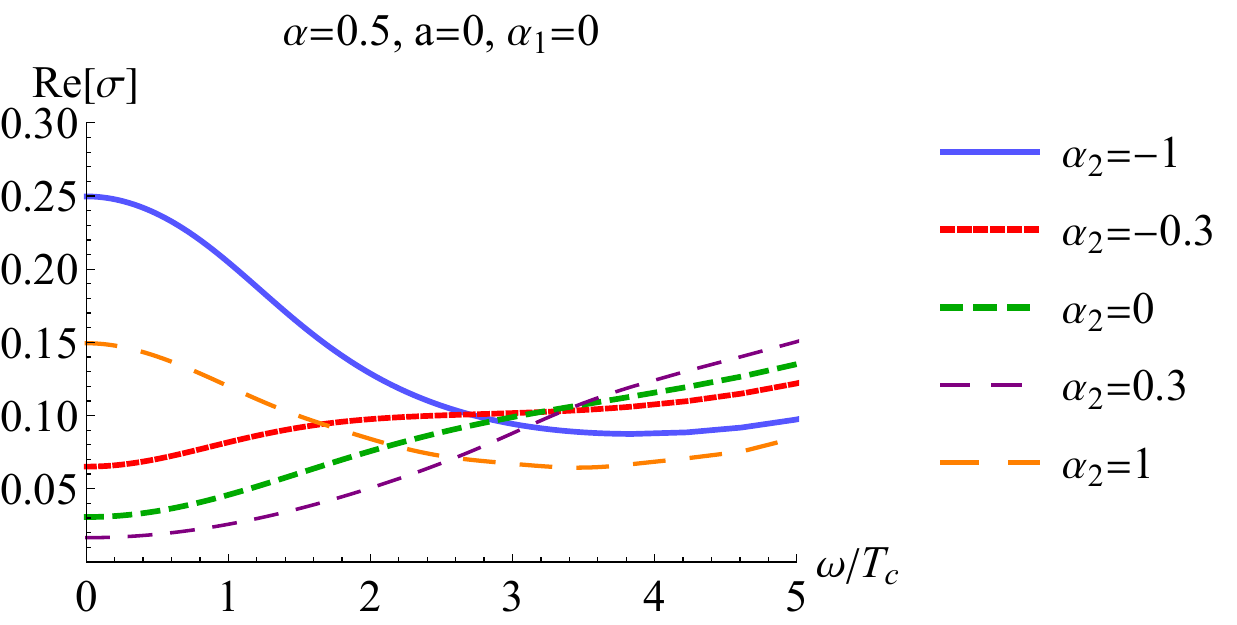}\\
        \caption{\label{na1listv1} The real part of conductivity as the function of the frequency at $T/T_{c}\approx0.2$. Left plot: different $\alpha_{1}$ for $\alpha=0.5$, $a=0$ and $\alpha_2=0$. Right plot: different $\alpha_{2}$ for $\alpha=0.5$, $a=0$ and $\alpha_1=0$.}}
\end{figure}
\begin{figure}
    \center{
        \includegraphics[scale=0.62]{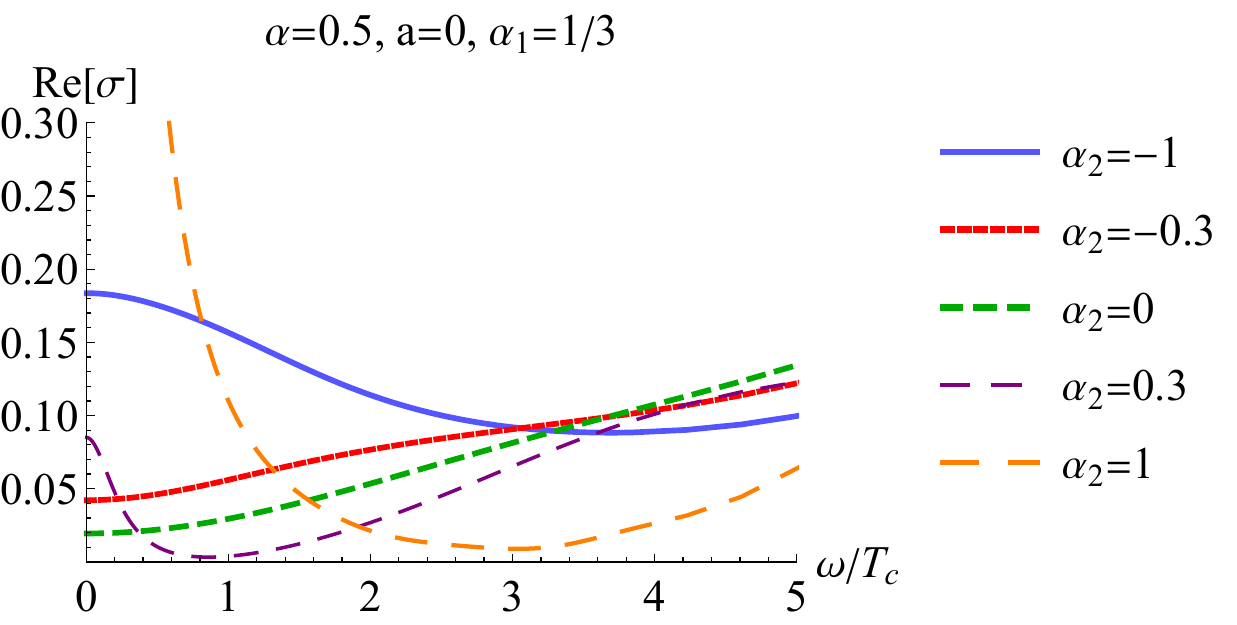}\ \hspace{0.5cm}
        \includegraphics[scale=0.62]{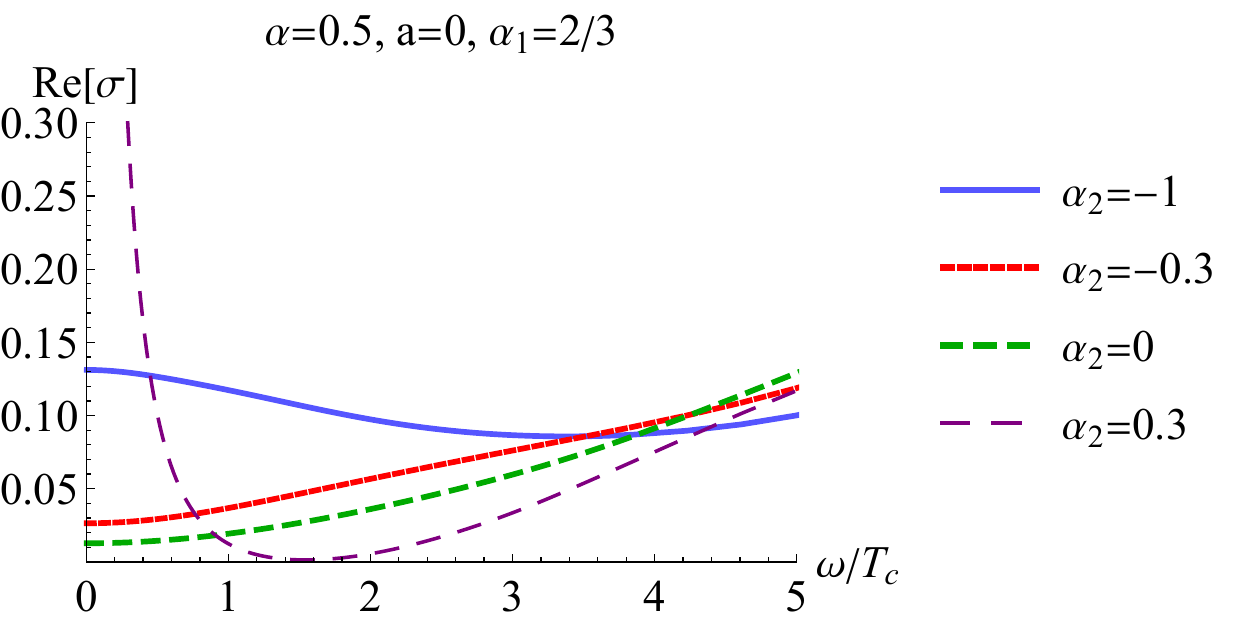}\
        \caption{\label{na1listv2} The real part of conductivity as the function of the frequency with various $\alpha_2$ for fixed $\alpha=0.5$ and $a=0$ at $T/T_{c}\approx0.2$. Left plot is for $\alpha_1=1/3$ and right plot for $\alpha_1=2/3$.
    }}
\end{figure}

Along that line, we also explore the impact of gauge-axion coupling parameters ($\alpha_{1}$ and $\alpha_{2}$) on the optical conductivity in the superconducting state. The results are exhibited in Fig.~\ref{na1listv1} and Fig.~\ref{na1listv2}. In order to avoid the effect of gauge parameter $a$ on the conductivity, here we fix $a=0$. In addition, we also set $\alpha=0.5$ without loss of generality. We summarize the main effects of the gauge-axion coupling as what follows:
\begin{itemize}
    \item Increasing the gauge-axion coupling $\alpha_1$, it is easier to drive the normal component of the electron fluid forming the superfluid component and the superconducting energy gap becomes evident (left plot in Fig.~\ref{na1listv1}). We infer that it is due to the remnant of the vortex response in the normal state resulting in more evident gap in the conductivity.
    This observation is consistent with that from high derivative holographic superconductor \cite{Liu:2020hhx,Wu:2017xki}.
    \item To see the effect from the coupling $\alpha_2$, which is a coupling among gauge field, axion fields and complex scalar field being responsible for condensation, we turn off $\alpha_{1}$. The low frequency behaviors of conductivity are exhibited in the right plot in Fig. \ref{na1listv1}. For negative $\alpha_{2}$, DC conductivity rises up with the absolute value of $\alpha_{2}$ increasing and then the conductivity forms a small peak at low frequency for large absolute value of $\alpha_{2}$. For positive $\alpha_2$, we see that with $\alpha_{2}$ increasing, DC conductivity goes down at first and then rises up exhibiting a non-linear change. When $\alpha_2$ rises up to a large value, the conductivity also sprouts up a small peak at low frequency. Based on these observations, it is obvious that the role $\alpha_{2}$ playing is completely different from that of $\alpha_{1}$. We attribute this difference to the effect of condensation field $\chi$. Furthermore, we would like to point out that the mechanism of this small peak emerging at low frequency deserving further pursuit.
    \item We further study the combined effects of the coupling $\alpha_{1}$ and $\alpha_{2}$, and also the competition between them. Fig.~\ref{na1listv2} shows the real part of conductivity with various $\alpha_{2}$ for fixed $\alpha_{1}=1/3$ (left plot) and $\alpha_{1}=2/3$ (right plot). It is clear that for negative $\alpha_{2}$, DC conductivity is suppressed with the increase of $\alpha_{1}$, which comes from the effect of remnant of vortex response.
    \item However, for positive $\alpha_2$, the effect of the coupling term $\alpha_{2}$ dominates over that of the coupling term $\alpha_{1}$. From Fig.~\ref{na1listv2}, we see that once the coupling terms $\alpha_1$ and $\alpha_{2}$ are both turned on, a peak emerges at low frequency. Recalling that for $\alpha_1=0$, a gap is evident at low frequency even for $\alpha_{2}=0.3$. With $\alpha_1$ increasing, the peak at low frequency becomes more evident.
\end{itemize}

\section{conclusion}\label{conclusion}

In this paper, we construct a holographic effective superconducting theory including some coupling terms among the gauge field, axion fields and the complex scalar field. In the normal state, this holographic effective superconducting theory reduces to the so-called $J$ model studied in Refs.~\cite{Li:2018vrz,Gouteraux:2016wxj}. We first study AC conductivity in the normal state, which is still absent as far as we know. In the normal state, though the gauge coupling term $\alpha_1\,{\rm Tr}[XF^{2}]$ has no effect on the background solution, it enters into the perturbative equations affecting AC conductivity. This term plays the role of the vortex exhibiting a dip in the low frequency conductivity. When the momentum dissipation is weak, there is a competition between the momentum dissipation and the gauge-axion coupling term. The gauge-axion coupling drives AC conductivity at low frequency away from the standard Drude peak when it is small and then a small but evident pronounced peak is observed at intermediate frequency, which indicates that the vortex response from gauge-axion coupling begin to emerge. As the gauge-axion coupling further increasing, an obvious spectral weight transfer from the low frequency to the intermediate frequency indicating that the vortex response dominates over the momentum dissipation. When both the gauge-axion coupling and the strength of momentum dissipation become large, a dip emerges. It is the result of a combination of the strength of the momentum dissipation and the gauge-axion coupling.

In the superconducting phase, we first study the properties of the condensation. Since the gauge-axion coupling term $J\,{\rm Tr}[XF^{2}]$ has no effect on the black hole solution, there are only two theoretical parameter considered in our model (the strength of momentum dissipation $\alpha$ and the gauge coupling parameters $a$) exerting their influences on the condensation. It is easy to find that with the parameter $\alpha$ increasing, the condensation becomes harder for fixed parameter $a$, while fixed parameter $\alpha$ with the parameter $a$ increasing, the condensation becomes easier. Such an opposite effect means a competitive relationship between $\alpha$ and $a$. Similarly, competitive behavior can be observed in superconducting conductivity.

Then we also study the properties of AC conductivity in the superconducting phase. We respectively explore the effects of coupling parameters ($\alpha$, $a$, $\alpha_{1}$, $\alpha_{2}$) in our model on the conductivity. First, we explore the effect of momentum dissipation on AC conductivity. The features of the conductivity with temperature from normal phase to superconducting phase resemble a two-fluid model as the standard holographic superconductor model \cite{Hartnoll:2008vx,Hartnoll:2008kx,Horowitz:2010gk}, and the holographic Q-lattice superconductor \cite{Ling:2014laa}, and the holographic superconductor from higher derivative theory \cite{Wu:2017xki,Liu:2020hhx}. Then for the influence of  the gauge coupling $a$, we find that $a$ plays a key role in driving the normal component of the electron fluid forming the superfluid component. Especially, for positive $a$, it is harder to implement the transition from the normal component of the electron fluid forming the superfluid component. In the end, we explore the effect of the gauge-axion coupling. We find that the coupling parameter $\alpha_{1}$ can drive the electron superfluid component forming which is similar with the effect of negative $a$. The increasing of $\alpha_{1}$ leads to a more evident gap in AC conductivity resulting from the remnant of the vortex response in the normal state. In addition, we find that the coupling of $\alpha_{2}$ plays a completely different role from that of $\alpha_1$, which is attributed to the effect of condensation field $\chi$.

\acknowledgments
We are very grateful to Guoyang Fu for helpful discussions and suggestions. J.-P. Wu would also like to thank Weijia Li, Peng Liu for previous collaboration or discussion. This work is supported by the Natural Science Foundation of China under Grant Nos.~11775036, 12147209 and 11975072, Fok Ying Tung Education Foundation under Grant No.~171006, the Liaoning Revitalization Talents Program (Grant No.~XLYC1905011), the National 111 Project of China (Grant No.~B16009), the Postgraduate Research \& Practice Innovation Program of Jiangsu Province (KYCX21\_3192), and Top Talent Support Program from Yangzhou University.

\bibliographystyle{style1}
\bibliography{Ref}
\end{document}